%% file: paper.tex
\documentclass[sigconf]{acmart}
\PassOptionsToPackage{hyphens}{url}
\usepackage[hyphens]{url}
\usepackage{hyperref} 
\hypersetup{pdfstartview=FitV,
  breaklinks=true, colorlinks=true, linkcolor=black, citecolor=black, urlcolor=blue
}
\usepackage{xspace}
\usepackage{enumitem}
\usepackage{dcolumn}
\usepackage{multirow}
\usepackage{subcaption}
\usepackage{microtype}
\usepackage{color}
\usepackage[final]{changes}
\usepackage{balance}

\graphicspath{{figures/}}

\newcommand{\yrp}{Yarrp\xspace}
\newcommand{\yrpsix}{Yarrp6\xspace}
\newcommand{\vfour}{IPv4\xspace}
\newcommand{\vsix}{IPv6\xspace}
\newcommand{\icmpsix}{ICMPv6\xspace}
\newcommand{\eg}{e.g.,\ }
\newcommand{\ie}{i.e.,\ }
\newcommand{\etal}{et al.\ }

\newcommand{\pt}[1]{(#1)}

\newcommand{\squishlist}{
 \begin{list}{${\bullet}$}
  { \setlength{\itemsep}{0pt}
     \setlength{\parsep}{3pt}
     \setlength{\topsep}{3pt}
     \setlength{\partopsep}{0pt}
     \setlength{\leftmargin}{1.5em}
     \setlength{\labelwidth}{1em}
     \setlength{\labelsep}{0.5em} } }
\newcommand{\squishend}{
  \end{list}  }

\copyrightyear{2018} 
\acmYear{2018} 
\setcopyright{usgovmixed}
\acmConference[IMC '18]{2018 Internet Measurement Conference}{October 31-November 2, 2018}{Boston, MA, USA}
\acmBooktitle{2018 Internet Measurement Conference (IMC '18), October 31-November 2, 2018, Boston, MA, USA}
\acmPrice{15.00}
\acmDOI{10.1145/3278532.3278559}
\acmISBN{978-1-4503-5619-0/18/10}

\definechangesauthor[color=orange]{2}

\begin{document}

\title{
In the IP of the Beholder: Strategies for Active IPv6 Topology Discovery
}

\author{Robert Beverly}
\orcid{0000-0002-5005-7350}
\affiliation{%
  \institution{Naval Postgraduate School}
}
\email{rbeverly@nps.edu}

\author{Ramakrishnan Durairajan}
\orcid{0000-0003-2859-5598}
\affiliation{%
  \institution{University of Oregon}
}
\email{ram@cs.uoregon.edu}

\author{David Plonka}
\orcid{0000-0003-0736-7297}
\affiliation{%
  \institution{Akamai Technologies}
}
\email{plonka@akamai.com}

\author{Justin P. Rohrer}
\orcid{0000-0002-3335-5612}
\affiliation{%
  \institution{Naval Postgraduate School}
}
\email{jprohrer@nps.edu}

\title[In the IP of the Beholder]{In the IP of the Beholder: Strategies for Active IPv6 Topology Discovery}
\renewcommand{\shortauthors}{R. Beverly et al.}

\begin{abstract}

Existing methods for active topology discovery within the \vsix Internet largely mirror
those of \vfour.  In light of the large and sparsely populated 
address space, in conjunction with aggressive \icmpsix rate limiting by
routers, this work develops a different approach to
Internet-wide \vsix topology mapping.  We adopt randomized
probing techniques in order to distribute probing load, minimize the
effects of rate limiting, and probe at higher rates.  Second, we
extensively analyze the efficiency and efficacy of various \vsix hitlists and target
generation methods when used for topology discovery, and synthesize
new target lists
based on our empirical results to provide both breadth (coverage across networks) and depth (to find
potential subnetting).  Employing our probing strategy, we discover 
more than 1.3M \vsix router interface \added{addresses} from a single
vantage point. 
\deleted[id=2]{-- 
an
order of magnitude more than produced by current state-of-the-art
mapping systems
that use hundreds of vantages.}
Finally, we \deleted{publicly} share our prober implementation, synthesized
target lists,
and discovered \vsix topology results.
\end{abstract}

\ccsdesc[500]{Networks~Network measurement}
\ccsdesc[300]{Networks~Topology analysis and generation}

\keywords{IPv6, Topology, Hitlists}

\maketitle

\input{intro}
\input{background}
\input{method}

\input{probing}
\input{results}

\input{subnet-discovery}

\input{subnet}
\input{discuss}

\section*{Acknowledgments}
\added{
We thank Eric Gaston for initial IPv6 Yarrp work, Matthew Luckie
and Emile Aben for early feedback, and Italo Cunha for shepherding.
KC Ng performed a trace campaign with a proprietary prober
for validation of~\yrpsix results by comparison.
Mike Monahan, Young Hyun, and Will van Gulik provided invaluable
infrastructure support.
DNSDB access supporting this research was generously granted by Farsight Security.
Views and conclusions are those of the
authors and should not be interpreted as representing the official
policies or position of the U.S.\ government.
}

\newpage
\clearpage
\balance
\urlstyle{same}
\bibliographystyle{ACM-Reference-Format}
\bibliography{yarrp}

\end{document}

%% file: intro.tex

\section{Introduction}\label{sec:intro}

As of \replaced{August}{May} 2018, about 23\% of Google's users access their services via
\vsix~\cite{GoogleIPv6Adoption}, while APNIC reports that
$\sim$\replaced{16}{15}k Autonomous
Systems (ASes) advertise \vsix prefixes~\cite{potaroo}. 
The number of \vsix routes in the BGP system has increased
from $\sim$5k in 2011 to more than \replaced{56}{48}k today~\cite{potaroo}, while native IPv6
adoption and traffic continues its exponential increase~\cite{CAZ+14}.
Similarly, a large content delivery network (CDN) observed 1.72B unique native
\vsix addresses for 576 million unique {\tt /64} prefixes on March 17,
2018~\cite{MAPRGActiveIPv6}. In a weeks' time, the number of \vsix~{\tt /64}s
covering active WWW clients approximated the total number of globally routed
\vfour unicast addresses ($\sim$2.5B). These examples
\replaced[id=2]{underscore the importance of \vsix today, and the}{suggest that \vsix
adoption is here, 
underscoring the acute} need for accurate \vsix topologies within the community.

Understanding the Internet's \vsix topology is important for
applications ranging from improved content distribution and traffic
optimization, to better address anonymization~\cite{DBLP:journals/corr/PlonkaB17}
and reputation~\cite{collins2007using, kalafut2010malicious}, to enhanced
network security~\cite{zhang2008ispy, syamkumar2016bigfoot}. Despite these
compelling applications, three challenges remain: \pt{i} an infeasibly large
address space that cannot be exhaustively scanned or uniformly sampled
effectively, \pt{ii} mandated and aggressive \icmpsix rate limiting in
routers~\cite{rfc4443}, and \pt{iii} unknown address allocation policies and
subnet structures. Note that the first two issues are inter-related: attempting
to increase coverage by probing more of the \vsix address space necessitates
faster probing rates. However, increasing the probing rate is
self-defeating as doing so triggers more rate limiting and, hence, fewer
discovered router interfaces and less
\replaced{representative}{accurate} topologies.

While decades of research have developed and refined active \vfour
topology discovery (\eg~\cite{spring2002measuring, claffy09,
donnet2005efficient, 1672318, Beverly:2010:PAI:1879141.1879162},
these
techniques do not address the aforementioned challenges unique to
\vsix.  Existing \vsix topology mapping systems are thus forced to
directly apply \vfour discovery tools and techniques. For example,
production CAIDA and RIPE \deleted{Atlas} traceroutes regularly
probe the {\tt ::1} address of every global IPv6 BGP
prefix~\cite{caida-topov6,ripeatlas}. Because of this very sparse sampling
\added{within the large \vsix address space}, the
completeness and quality of the resulting logical topologies
\replaced{is}{are}
unknown.

In this work, we seek to advance the state-of-the-art in Internet-wide
\vsix active topology mapping.  Our methodology tackles two
fundamental aspects of the problem: \added{\pt{i}} which addresses to
target; and \added{\pt{ii}}
how to probe.

First, we amass the largest collection of \vsix target
addresses/seeds \added{currently available} from a variety of sources (\eg BGP, DNS, CDNs
\cite{pam2017-nxdomain, rapid7, farsightPassiveDNS,
gasser2016ipv6scanning, DBLP:journals/corr/PlonkaB17})
as well as generated seeds (\eg 
6Gen~\cite{Murdock:2017:TGI:3131365.3131405}). We employ a three-step process
to synthesize 12.4M target addresses \added{specifically crafted to
\replaced[id=2]{promote}{enhance} topology discovery}. 
Next, we
perform \replaced{a total of 45.8M traces}{45.8M traces, in total,} from
three vantages: two US universities and one EU network.
We investigate different target selection methods and
 parameters (\eg maximum TTL, protocol, probing speed, etc.) to
\added{find those that} elicit the most \vsix
topological information.  We show that our methodology provides 
breadth across networks, depth to discover subnetting, and speed.
Employing our probing strategy, we discover
$>$1.3M \vsix router interface \added{addresses} from a single vantage point in
a single day -- an order of magnitude more than produced by current state-of-the-art
mapping systems
that use hundreds of vantages in the same period.
Our primary contributions thus include:
\squishlist
 \item Evaluation of various
       means to synthesize target addresses from seven different input \replaced{seed sets}{seeds}.
 \item Quantification of random probing to maintain high 
       rates while avoiding \icmpsix rate limiting.
 \item Characterization of target list power \replaced{to yield topological results}{and the resulting topologies}.
 \item \vsix subnet discovery as a case study of topological inference.
 \item \deleted{Public} \vsix \added{router interface-level} topology \replaced{results}{maps}, our synthesized target
       lists, and our prober implementation \cite{topodownload}.
\squishend

\added{The remainder of this paper is organized as follows.
Section~\ref{sec:background} provides background on \deleted[id=2]{the current
state-of-the-art in} \vsix topology mapping, while
Section~\ref{sec:method} characterizes the \deleted[id=2]{various starting} seed lists
we utilize and describes our method for synthesizing an \vsix target
set. \deleted[id=2]{for topology probing.}  We describe our probing
\deleted[id=2]{technique} in
Section~\ref{sec:method:probing}, and provide results from an
Internet-wide probing campaign in Section~\ref{sec:results}.
Section~\ref{sec:subnetdiscovery} analyzes the results to infer likely
subnetting and better understand provider provisioning.
\replaced{W}{Finally, w}e conclude in Section~\ref{sec:discuss} with a discussion of
the security and privacy implications of our work, as well as
suggestions for continued research.}

%% file: background.tex

\section{Background and Related Work}\label{sec:background}

\added{
{\bf \vfour topology.} 
Since the seminal \added[id=2]{Mercator}~\cite{govindan2000heuristics} and Rocketfuel work~\cite{spring2002measuring}, the
measurement community has made continual progress toward improving
\vfour topology discovery and inference (see \eg \cite{claffy09,
1672318, Beverly:2010:PAI:1879141.1879162} and references therein).  
Related to our initiative to
design a scalable active topology mapping system,
Doubletree~\cite{donnet2005efficient} begins tracing from a midpoint
\replaced[id=2]{on the path}{in the graph} and halts when a trace reaches a known, previously
probed portion of the path.  We examine Doubletree's performance for
\vsix in \deleted[id=2]{more detail in }Section \ref{sec:probing:tuning}.
}

\added{Recently, Beverly introduced \yrp, a randomized high-speed \vfour topology
prober~\cite{imc16yarrp}.  In contrast to traditional traceroute
techniques intended to probe individual paths, \yrp
spreads topology probes across the network rather than probing the path
to each individual destination sequentially.  It does this by
randomly permuting the space of destination targets and TTLs,
thereby attempting to avoid
overloading any single router or path.  In addition, \yrp is
stateless and can recover the necessary information to match 
replies via carefully crafted probes.  
These properties allow \yrp to perform high-speed topology mapping.
For instance, \yrp has been run at $100kpps$ to
discover more than $400k$ unique \vfour router addresses in
approximately 23 minutes~\cite{imc16yarrp}.  
}

\added{Despite the significant body of prior work on \vfour topology,
\vsix has at least two fundamental differences: a much larger address
space (that is sparsely populated) and \replaced{aggressive}{mandated} \icmpsix rate limiting~\cite{rfc4443, v6RouterReqs}.
We explore new techniques to accommodate both of these properties in
this work.}

{\bf \vsix topology.} 
Dhamdhere \etal studied the evolution of \vsix topology at
the AS-level using passive BGP data and found that fewer
than 50\% of AS-level paths in 2012 were identical between \vfour and \vsix, while a
single AS (Hurricane Electric) \replaced[id=2]{appeared in 20-95\% of
observed \vsix AS paths}{was dominant in the
topology}~\cite{dhamdhere2012}.  Similarly, Czyz \etal examined BGP
tables in 2013 and, while they found only 19\% of ASes supporting \vsix,
\deleted[id=2]{as compared to \vfour,} a k-core analysis showed that these ASes were
well-connected large networks with high centrality~\cite{CAZ+14}.
Complementary to these efforts, we take an in-depth look at the \vsix
topology, starting from an interface-level perspective, 
after a period of sustained growth, via 
\emph{active} probing.

\deleted{Prior \vsix topology work has largely avoided active probing due to
the sheer size of the address space and the sparsity of infrastructure
within it.}
Presently, two production measurement platforms
continually perform active \vsix topology mapping: CAIDA's
Ark~\cite{caida-topov6} and RIPE Atlas~\cite{ripeatlas}.  These
systems send Paris~\cite{augustin2006avoiding} traceroute probes toward the {\tt ::1} address in each \vsix prefix
present in the global BGP table.  Ark also probes a {\em random} address in each prefix.  A central finding of our work is
that using BGP prefixes alone to guide target selection works well to
capture topological breadth, but not depth, \ie it does not discover
subnetting. 

Rohrer \etal uniformly
sampled the \vsix Internet \added{in 2015} by tracerouting
to an address in every /48 prefix that makes up the routed IPv6 address space~\cite{v6exhaust-ic16}.  Their study issued
ten times as many traces as our work, yet found an order
of magnitude fewer interfaces -- \deleted{again} demonstrating the necessity
to perform finer-grained probing within some prefixes to discover
extant subnetting and the routers supporting the subnets.


More broadly, prior studies of \vsix
topology use traditional tools including 
{\tt traceroute6}~\cite{DBLP:conf/pam/Malone08} and {\tt
scamper}~\cite{luckie10scamper}.
Gaston was the first to explore higher-rate \vsix active topology
probing~\cite{gaston17} via randomization, and demonstrated the ability
to capture roughly an equivalent amount of topological information as
collected via CAIDA's Ark system.
However, Gaston's study did not examine the
critical problem of target selection and rate-limiting, and did not
explore the ability to utilize high-speed probing to discover a
larger swath of the topology. Alvarez \etal developed and evaluated
methods to deal with \icmpsix rate-limiting problems,
but in a stateful, proprietary prober~\cite{MAPRGRateLimitingIPv6}.


\added{Separate to interface-level topology discovery is
\deleted[id=2]{the task of}
finding \replaced{aliases}{aliased interfaces}, \ie determining those interfaces that
belong to the same physical router, and \deleted[id=2]{then } creating router-level
graphs.  Luckie \etal developed speedtrap, the first Internet-scale
\vsix alias resolution technique~\cite{imc13speedtrap}, now used to
produce router-level graphs of the \vsix \added[id=2]{Internet} as part of CAIDA's
ITDK~\cite{caida-itdk}.  In this work, we focus on \vsix interface
address discovery, and do not perform alias resolution.  However, as alias
resolution takes interface addresses as input,
\replaced[id=2]{the}{our} ability to discover \deleted[id=2]{more}
\vsix interfaces \replaced[id=2]{is directly beneficial}{benefits
alias resolution}~\cite{ramaUAv6}.}

{\bf \vsix hitlists.} \replaced{Catalogs of}{Cataloging} active \vsix addresses in the Internet, commonly
known as \replaced{{\em hitlists}}{{\em a hitlist}}, \replaced{have}{has} been of interest to the measurement community over
the last decade. Notable efforts leverage active (\eg random probing of
{\tt ::1}~\cite{caida-topov6}, exhaustive probing of
\replaced[id=2]{{\tt ::1}}{one} address in each {\tt
/48} in all advertised {\tt /32}s~\cite{v6exhaust-ic16}, and reverse DNS
zone walking~\cite{pam2017-nxdomain,BorgolteSP2018}) and passive techniques (\eg from BGP
updates~\cite{dhamdhere2012, potaroo} and from traffic
captures~\cite{dainotti2013estimating}) as well as a number of passive and
active sources~\cite{gasser2016ipv6scanning}.  \added{Similar to our
effort, Gasser \etal observe the propensity of \vsix hitlists to
contain clusters of addresses~\cite{imc18gasser}.  While Gasser provides a new,
aggregate \vsix hitlist that also removes aliased prefixes, they do
not study how to adapt the hitlist to \vsix topology discovery.}
We provide details of the
specific hitlists we utilize in Section \ref{sec:tas}.

{\bf \vsix addresses and deployments.} Many issues pertinent to our work
-- including
\added[id=2]{address} seed sources, target selection, and probing techniques -- reflect
current understanding of \vsix network
\replaced{reconnaissance}{reconnsaissance}~\cite{rfc7707}.
Malone analyzes different aspects of three \vsix
address datasets and presents an analysis technique to learn about their
deployments and usages~\cite{DBLP:conf/pam/Malone08}. Czyz 
\etal\cite{czyz2016} studied 520k and 25k dual-stacked servers and routers respectively and
compared the security policies of \vfour and \vsix deployments. 
To assess the lifetime and density of \vsix addresses,
Plonka and Berger
developed Multi-Resolution Aggregate plots and classified 
billions of addresses spatially and temporally~\cite{Plonka:2015:TSC:2815675.2815678}.
A recursive algorithm to discover and extract \vsix addressing patterns is
discussed in~\cite{ullrich2015reconnaissance}. Similarly, Foremski \etal
proposed the Entropy/IP system to uncover the structure of \vsix addresses
using machine learning~\cite{Foremski:2016:EUS:2987443.2987445}. In a \replaced{like}{similar}
vein, Murdock \etal designed ``6Gen'' target generation ~\cite{Murdock:2017:TGI:3131365.3131405}. 
6Gen exploits address locality:
discovery of new targets happens closer to highly dense ranges. In a nybble, targets are
generated either based on a specific range (\eg {\tt
2::[1-4]:0}) or wildcard (\eg {\tt 2::?:0}); the former is
  {\em tight} clustering while the latter is referred as {\em
loose}.

%% file: method.tex

\begin{figure}
 \centering
 \resizebox{1.0\columnwidth}{!}{\includegraphics{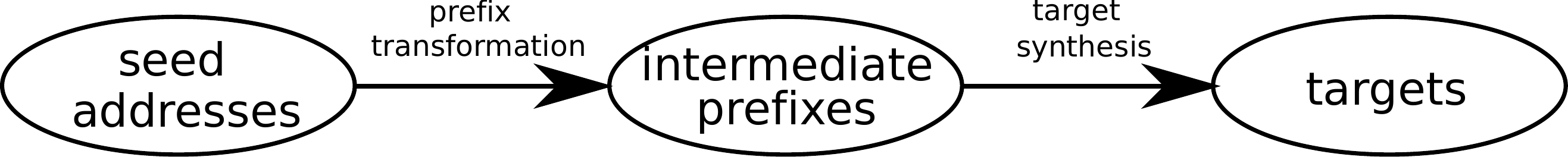}}
 \vspace{-5mm}
 \caption{\vsix topology target generation: addresses from various
 \emph{seeds} are transformed into \emph{intermediate
 prefixes} which are then synthesized into \emph{targets}.}
 \label{fig:targetselect}
 \vspace{-4mm}
\end{figure}

\section{Target Selection}
\label{sec:method}

Our methodology tackles two primary challenges: \pt{i}
selecting targets from the large and sparsely populated \vsix address
space; and \pt{ii} \replaced{effectively probing}{effectively sourcing
probes}.
This section considers target selection.

\newcolumntype{d}[1]{D{.}{.}{#1}}
\input{seedListProperties_table_scaled}

\subsection{Target Generation}\label{sec:tas}

We employ a three step process, outlined in
Figure~\ref{fig:targetselect}, to prepare a set of {\em targets},
based on a set of {\em intermediate prefixes}, which 
are synthesized from {\em seeds}.  The targets are
IP addresses to be used as destinations for TTL-limited probes
emitted from a {\em vantage}.  \added{When discussing \vsix
addresses, we adopt the standard vernacular of a subnet {\em prefix}
to denote the high-order bits and the {\em interface 
identifier} (IID) to denote the least significant 64 bits~\cite{rfc4291}.}

{\bf Step 1: Seed Sourcing.} A list of {\em seeds} is obtained from a {\em source}.
Our seeds are either IP prefixes (base address and length) or \vsix addresses 
(base address with implicit 128\replaced{bit}{b} length), which are used as hints
to select interesting areas of the address space that help define probe
destinations.  

{\bf Step 2: Prefix Transformation.} One or more {\em transformations} may be applied to the seeds
to yield a set of {\em intermediate prefixes}. Depending on
the transformation method, the resulting set might have 
the same or fewer number of prefixes than the seed list to which it
was applied. 
The prefix transformations \replaced{we use are:}{used
in this study are as follows:}

\squishlist
\item {\bf k{\em n}}: Perform kIP aggregation-based address
anomymization with parameters: $w=14$ (window days), $i=1$ (interval
hours), $k=n$ (simultaneously-assigned /64 prefixes), $p=50$ (50th percentile of intervals).~\cite{DBLP:journals/corr/PlonkaB17}, \eg {\bf k32} and {\bf k256}~\cite{DBLP:journals/corr/PlonkaB17}.
\item {\bf z{\em n}}: Extend input prefixes with length $<n$ to /n, 
       \ie base address set to zeroes after the {\em n}th bit; 
       aggregate input prefixes having length $>n$ to /n prefixes.
\squishend

{\bf Step 3: Target Synthesis.} Lastly, a {\em synthesis} method
takes the intermediate prefixes as input and yields a set of target
addresses.  
The target synthesis methods used in
this study are as follows:

\squishlist
\item {\bf lowbyte1}:
bitwise OR prefix base address with IID value \\
{\tt :0000:0000:0000:0001}.
\item {\bf fixediid}:
bitwise OR prefix base address with IID value \\
{\tt :1234:5678:1234:5678}.
\squishend

\noindent For each, we remove duplicate addresses within
the set.  The target address sets are then ready to be employed in
a probe {\em campaign}. 


\subsection{Seeds}


\noindent The seed sources used in this study are as follows:
\squishlist
\item {\bf caida}: The set of probe targets selected by CAIDA, based on
 BGP-advertised \vsix prefixes of size /48 or larger, \ie prefixes
 with length of
 at most 48 bits~\cite{caida-topov6}.

\item {\bf fiebig}: The set of \vsix addresses gleaned from walking the {\tt
  ip6.arpa} zones in the DNS~\cite{pam2017-nxdomain}.

\item {\bf fdns\_any}: A set of \vsix addresses found in DNS answers in response
  to forward DNS ANY queries performed by Rapid7's Project
  \replaced{Sonar~\cite{rapid7}}{Sonar.\cite{rapid7}}.

\item {\bf dnsdb}: A set of \vsix addresses found in DNS answers passively
  observed in AAAA DNS query responses by Farsight Security's Farsight Passive
  DNS project\replaced[id=2]{,}{and} anonymized and imported into Farsight
  DNSDB~\cite{farsightPassiveDNS}.  We queried DNSDB \added[id=2]{for} all \vsix records observed
  between 15 Feb and 28 Apr 2018 in the covering set of all advertised BGP \vsix
  prefixes (as reported by RouteViews~\cite{routeviews} on 20 Apr 2018).

\item {\bf cdn}: \replaced{A set of {\vsix} prefixes, having various lengths, \added{that are {\em anonymized aggregates}~\cite{DBLP:journals/corr/PlonkaB17}} covering WWW client \vsix addresses believed}{A set of \vsix /64 prefixes that covered the set \vsix
WWW client addresses thought} to be SLAAC temporary privacy addresses \added{(the most common type of WWW client addresses~\cite{Plonka:2015:TSC:2815675.2815678})},
as observed by a large Content Delivery Network (CDN) across 14
days, February 18, 2018 through March 3, 2018 (UTC).

\item {\bf 6gen}: A set of \vsix address synthesized using
\added[id=2]{the} 6Gen tool in loose
clustering mode~\cite{Murdock:2017:TGI:3131365.3131405}. The input to the tool
is a combination of \vsix destinations probed and new interfaces found from
probing those destinations by CAIDA on March 06, 2018. 

\item {\bf tum}: A combined set of \vsix addresses created by combining multiple
existing address sets including some that we also use individually (fdns\_any) and
a number of others (openipmap, ct, caida-dnsnames, alexa-country, traceroute, 
and traceoute-v6-builtin).  Some of these subsets are documented in~\cite{gasser2016ipv6scanning}
and others are undocumented.  Each subset is packaged separately, and we noted some anomalies
(such as zero-size of very small files for recent dates) leaving room for interpretation as
to what would best represent the TUM dataset as a whole.  With this in mind we took the 
most recent and largest file for each subset, which are shown in Table~\ref{tab:tum}.

\begin{table}[t]
 \small
 \caption{TUM \added{Seed} Subsets} \label{tab:tum}
 \vspace{-3mm}
 \centering
 \begin{tabular}{|r|r|}
   \hline
   \multicolumn{1}{|c|}{\textbf{Filename}} & \multicolumn{1}{c|}{\textbf{\# Addresses}} \\
   \hline\hline
    {\tt alexa-country-2018-04-23.csv} & 1,634 \\
    \hline
    {\tt caida-dnsnames-2018-01-17.csv} & 1,268,982  \\
    \hline
    {\tt caida-dnsnames-2018-04-25.csv} & 16,507  \\
    \hline
    {\tt ct-2018-04-27.csv} & 21,983,387 \\
    \hline
    {\tt openipmap-2018-04-23.txt} & 11,149 \\
    \hline
    {\tt rapid7-dnsany-2018-03-30.csv} & 24,959,477 \\
    \hline
    {\tt rapid7-dnsany-2018-03-31.csv} & 14,434,117 \\
    \hline
    {\tt traceroute-2018-04-24.txt} & 128,398 \\
    \hline
    {\tt traceroute-v6-builtin-2018-02-25.txt} & 112,754 \\
    \hline
    All zone files from {\tt 2018-04-27} & 17,173,243 \\
    \hline
    \textbf{Total} & 80,089,648 \\
    \hline
    \textbf{Total Unique} & \textbf{5,599,313} \\
    \hline
 \end{tabular}
 \vspace{-4mm}
\end{table}


\squishend


\noindent We examine several features of our seed lists in Table~\ref{tab:seeds}.  At a glance we note
the wide variance in size (105k addresses to 25M).  The seeds are derived using
widely varying \replaced{methodologies}{methodology}, which we hope makes them complementary (not redundant).
\deleted[id=2]{We specify the date of acquisition; in cases where the seeds correspond to a range of dates,
just the end date is specified in the table to conserve space.  The full range is given in
Step 1.}
We \added{also} perform \replaced{a}{of} \replaced{simple}{rough} classification of the IIDs in order to inform our choice of 
target IID, as well as enable correlation with result sets.

We classified our seed addresses using the {\tt addr6}
tool~\cite{ipv6toolkit}. This tool examines each address, looking
for patterns, such as whether the IID, \eg
\pt{i} may be an EUI-64 IID with an embedded MAC address,
\deleted[id=2]{(``IEEE-based''),}
\pt{ii} has a run of zeroes followed only by a low number (``lowbyte''), or
\pt{iii} has no discernible pattern (``randomized'', essentially meaning
unrecognized). In Table~\ref{tab:seeds}, the percentages reflect the proportion within each seed list (row).

\deleted[id=2]{A few notes on particular seed lists are worth mentioning.} The CDN \replaced{clients'}{Clients} individual \vsix addresses
\replaced{were not provided to the authors in order to maintain privacy; instead, the authors received and utilized anonymized prefixes that were}{are unavailable to the authors due to privacy concerns, instead prefixes are provided,} generated 
using the kIP aggregation approach \replaced{as mentioned}{described} in
Section~\ref{sec:tas}.  While the number of addresses
aggregated is \replaced{undisclosed}{unknown}, there were 421,807 aggregates \added{(prefixes)} in the CDN-k256 set and 3,445,329
\deleted{aggregates} in the CDN-k32 set.  We treat the first six seed lists in Table~\ref{tab:seeds} as 
independent.  We create our own combined list by joining those six lists.  The TUM list is also a 
collection~\cite{gasser2016ipv6scanning}, which includes CAIDA and FDNS subsets, and therefore
is not independent of the first six lists.  Lastly, we randomly generate 26.5M targets in BGP routed \vsix 
address space as a control seed list.

\subsection{Selecting Transformations}
\label{sec:method:xfrm}


{\bf Prefix transformation granularity.} 
Next, we seek to understand the influence of aggregation granularity
when performing prefix transformation.  Many of our seed input
datasets contain multiple \vsix addresses within the same /64 prefix --
this is a natural consequence of their intended use as hitlists for
\vsix host discovery, rather than \vsix router discovery.  Intuitively,
assuming that /64 prefixes are \deleted{frequently }allocated to \replaced{LANs or Internet service subscribers as}{customers and represent} the smallest common \vsix subnets, we would not \added{typically} expect traceroutes to
multiple addresses within the same /64 to yield different topologies.

To characterize the relationship to the discovered topology, and the
probing required when using \replaced{our {\bf z{\em n}} transformation}{\bf z{\em n}}, we \added{mounted a trial campaign that} probed the fdns\_any
dataset for varying values of $n$.  Table~\ref{tab:agglevel} shows
that {\bf z64} requires more than eight times as many probes as {\bf
z40}, but discovers three times as many unique interfaces.  More
importantly, we consider the number of interfaces that are discovered
exclusively as a result of using a particular \replaced{transformation}{aggregation} level.
Although {\bf z64} requires significant probing, there are more than
27k interfaces that \added[id=2]{were} only discovered \added{when} using this \replaced{transformation}{aggregation} level.  Finally,
we examine the number of non-``Time Exceeded'' responses from each of the
\deleted{aggregation} levels.  The \replaced{fraction}{number} of \replaced{other}{other-} \icmpsix responses per probe is
0.012, 0.029, 0.032, and 0.041 for $n=40,48,56,64$, respectively. Thus, after
normalizing to the number of probes, the $n=64$ \replaced{transformation}{aggregation} level has
the effect of producing a higher rate of non-``Time exceeded'' responses,
suggesting that these probes are reaching \replaced{deeper}{further} into \deleted{the} network\added{s}.

\input{agglevel_table_scaled}


{\bf Target synthesis.} 
The lower 64 bits of an \vsix address denote the interface identifier \added{(IID)}
\cite{rfc4291}.  Given a candidate prefix, we must select \replaced{an IID}{the host
identifier} within that prefix to probe\deleted{ toward}.  Natural candidates are
the \added{IID} \texttt{::1}, a random \replaced{IID}{identifier}, \replaced{an IID occurring in}{a known host identifier from}
the input seed list, or a \replaced{(optionally fixed)}{fixed} pseudo-random \replaced{identifier}{IID}.  In addition
to understanding whether this choice has an impact on topological
discovery, we further wish to expose the extent to which different
identifiers elicit non-``Time Exceeded'' messages as a metric of impact on
hosts.  We therefore mounted \replaced{two trial campaigns}{a campaign} using the {\tt cdn-k256} prefixes\added{, the {\bf z64} transformation,}
and both the \pt{a} {\bf lowbyte1} and \pt{b} {\bf fixediid} host identifiers for
target synthesis.  Table~\ref{tab:hostid} shows the distribution of
\icmpsix Time Exceeded and Destination Unreachable responses received as a result of the campaigns.

\begin{table}[t]
 \small
 \caption{\replaced{\icmpsix Trial Results by IID}{Host Identifier}}\label{tab:hostid}
 \vspace{-3mm}
 \centering
 \begin{tabular}{|c|c|c|c|}
   \hline
   \multirow{2}{*}{\textbf{type/code}} & \multicolumn{2}{c|}{\textbf{CDN-k256 z64}} & \textbf{Fiebig} \\
   \cline{2-3}
   & {\bf lowbyte1} & {\bf fixediid}  & {\bf known} \\
   \hline\hline
    Time Exceeded               & 98.1\% & 98.1\% & 95.8\% \\\hline
    no route to destination     & 0.7\% & 0.7\%   & 0.6\%  \\\hline
    administratively prohibited  & 0.6\% & 0.6\%   & 0.4\%  \\\hline
    address unreachable         & 0.3\% & 0.4\%   & 0.7\%  \\\hline
    port unreachable            & 0.1\% & 0.0\%   & 2.3\%  \\\hline
    reject route to destination & 0.1\% & 0.2\%   & 0.2\%  \\\hline
 \end{tabular}
 \vspace{-3mm}
\end{table}

First, we find that more than 98\% of responses are \icmpsix time
exceeded messages as we expect.  We see only negligible differences
between {\bf lowbyte1} and {\bf fixediid}; {\bf lowbyte1} produces
five times as many \icmpsix port unreachable responses, but the total
number of such responses is still very small.  \added{Second,} to understand the use
of a known address within the prefix, we further compare against
probes toward known addresses within the Fiebig seed list \added{(not shown)}.  In
contrast to probing the \replaced{{\tt ::1} or fixed, these}{base or fixed identifier,} port unreachable
messages constitute 2.3\% of the distribution for the known address
probing, suggesting that our probing is reaching \deleted{these }end hosts.

Because we observe only minimal impact on discovery between fixed
\replaced{IID and the {\tt ::1} IID}{interface identifiers and the
base address} \added[id=2]{(a response rate difference of less than
2\%)}, we choose to use the fixed
identifier for the remainder of the experiments in order to minimize
any potential \replaced{impacts from probes reaching end hosts}{impact on end hosts}.

\input{targets}

%% file: seedListProperties_table_scaled.tex

\begin{table*}[t]
  \centering
  \small
  \caption{Seed List Properties}\label{tab:seeds}
  \vspace{-3mm}
  \begin{tabular}{|c|c|c||r|r|d{2.3}|r|d{2.3}|r|d{2.3}|}
    \hline
    \multirow{2}{*}{\textbf{Name}} & \multirow{2}{*}{\textbf{Method}} & \textbf{Date} & \multicolumn{1}{c|}{\multirow{2}{*}{\textbf{\# Addrs}}} & \multicolumn{6}{c|}{\textbf{Interface Identifiers (IIDs)}} \\
    \cline{5-10}
    &  & yyyy/mm/dd &  & \multicolumn{2}{c|}{\textbf{Random}} & \multicolumn{2}{c|}{\textbf{LowByte}} & \multicolumn{2}{c|}{\textbf{EUI-64}} \\
    \hline\hline
    \textbf{CAIDA}~\cite{caida-topov6} & BGP-derived & {2018/05/09} & 105.2k & 53.7k & 51.02\% & 51.5k & 48.98\% & {0} & 0.00\% \\
    \hline
    \textbf{DNSDB}~\cite{farsightPassiveDNS} & Passive DNS & {2018/02/15} -- {2018/04/28} & 5.4M & 1.3M & 24.43\% & 2.2M & 41.27\% & 146.5k & 2.74\% \\
    \hline
    \textbf{Fiebig}~\cite{pam2017-nxdomain} & Reverse DNS & {2018/03/27} & 11.7M & 4.2M & 35.94\% & 3.2M & 27.54\% & 275.4k & 2.35\% \\
    \hline
    \textbf{FDNS}~\cite{rapid7} & Fwd. DNS & {2018/04/27} & 24.8M & 3.3M & 13.12\% & 7.0M & 28.20\% & 236.8k & 0.95\% \\
    \hline
    \multirow{2}{*}{\textbf{CDN Clients}\cite{DBLP:journals/corr/PlonkaB17}} & {\em k}IP anonymization: $k=256$ & {2018/02/18} -- {2018/03/03} & N/A & All & 100.00\% & {0} & 0.00\% & {0} & 0.00\% \\
    \cline{2-10}
    & {\em k}IP anonymization: $k=32$ & {2018/02/18} -- {2018/03/03} & N/A & All & 100.00\% & {0} & 0.00\% & {0} & 0.00\% \\
    \hline
    \textbf{6gen}~\cite{Murdock:2017:TGI:3131365.3131405} & Generative
    & \replaced{2018/03/13}{2018/02/13} & 4.9M & 4.4M & 89.61\% & 389.2k & 7.93\% & 17.1k & 0.35\% \\
    \hline
    \textbf{Combined} & Join Sets & varies & 50.8M & 13.2M & 28.19\% & 12.9M & 27.45\% & 675.8k & 1.44\% \\
    \hline\hline
    \textbf{TUM}~\cite{gasser2016ipv6scanning} & Collection & varies & 5.6M & 2.3M & 44.62\% & 1.2M & 23.54\% & 604.0k & 11.79\% \\
    \hline
    \textbf{Random} & Random & {2018/05/23} & 26.5M & {0} & 0.0\% & 95.8k & 0.36\% & {0} & 0.0\% \\
    \hline
  \end{tabular}
  \vspace{-3mm}
\end{table*}

%% file: agglevel_table_scaled.tex

\begin{table}[t]
 \small
 \caption{\replaced{\icmpsix Trial Results by Transformation}{Aggregation Level}}\label{tab:agglevel}
 \vspace{-3mm}
 \centering
 \begin{tabular}{|c|c|c|c|c|}
   \hline
   {\bf z{\em n}} & Probes & Other \icmpsix & Addrs & Excl\ Addrs\\
   \hline\hline
   /40  & 1.4M  & 17.5k  & 27.0k & 158     \\ \hline
   /48  & 3.6M  & 105.8k & 45.5k & 321     \\ \hline
   /56  & 6.1M  & 194.8k & 60.5k & 1.1k  \\ \hline
   /64  & 11.8M & 486.8k & 85.5k & 27.2k \\ \hline
 \end{tabular}
 \vspace{-5mm}
\end{table}

%% file: targets.tex


\begin{figure}[tb]
  \centering
    \includegraphics[width=\linewidth]{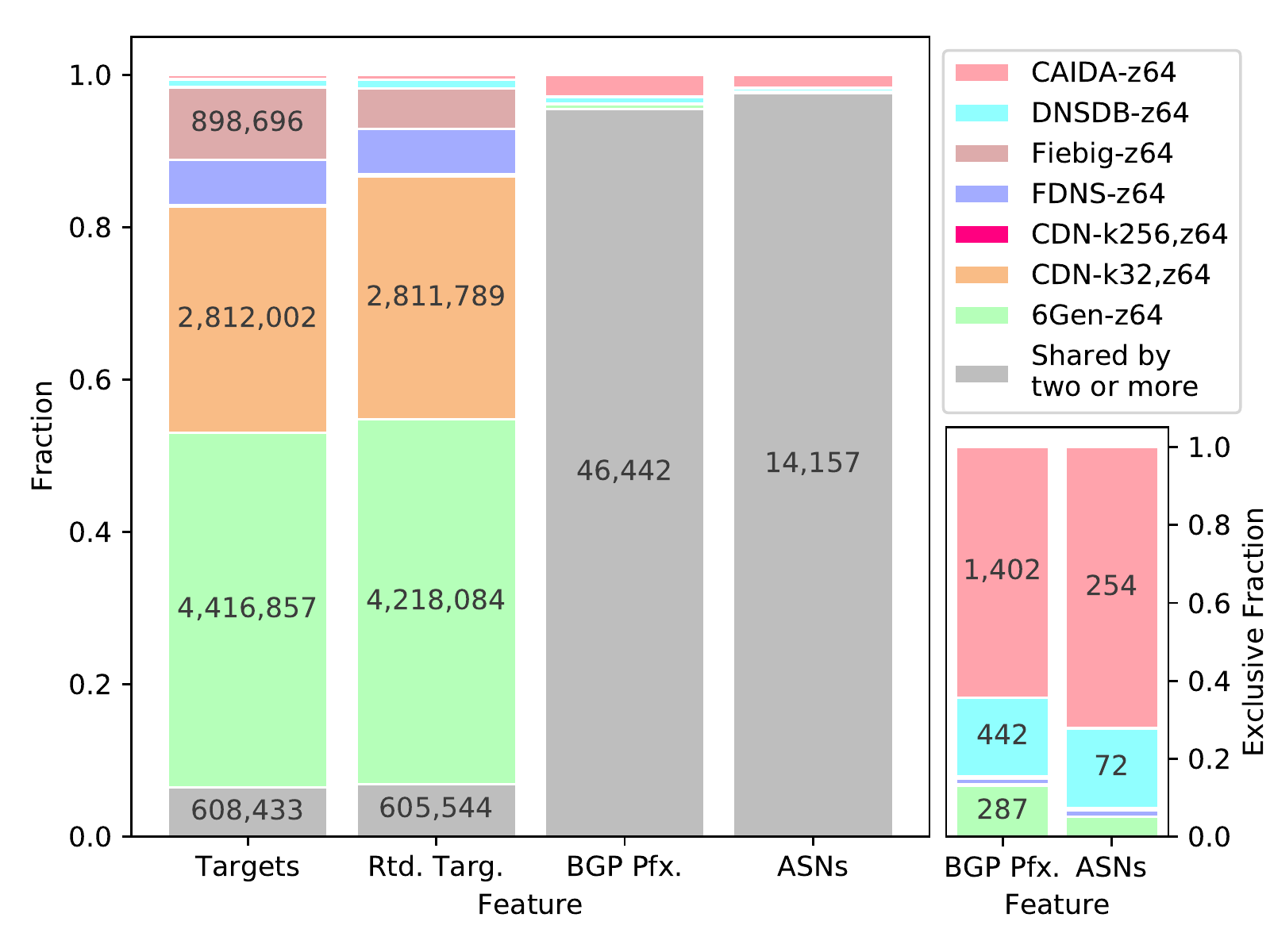}
    \vspace{-7mm}
    \caption{Features contributed by each target set}
    \label{fig:targ-stack}
    \vspace{-4mm}
\end{figure}


\begin{figure}[tb]
\centering
  \begin{subfigure}{0.42\textwidth}
  \centering
  \includegraphics[width=\linewidth]{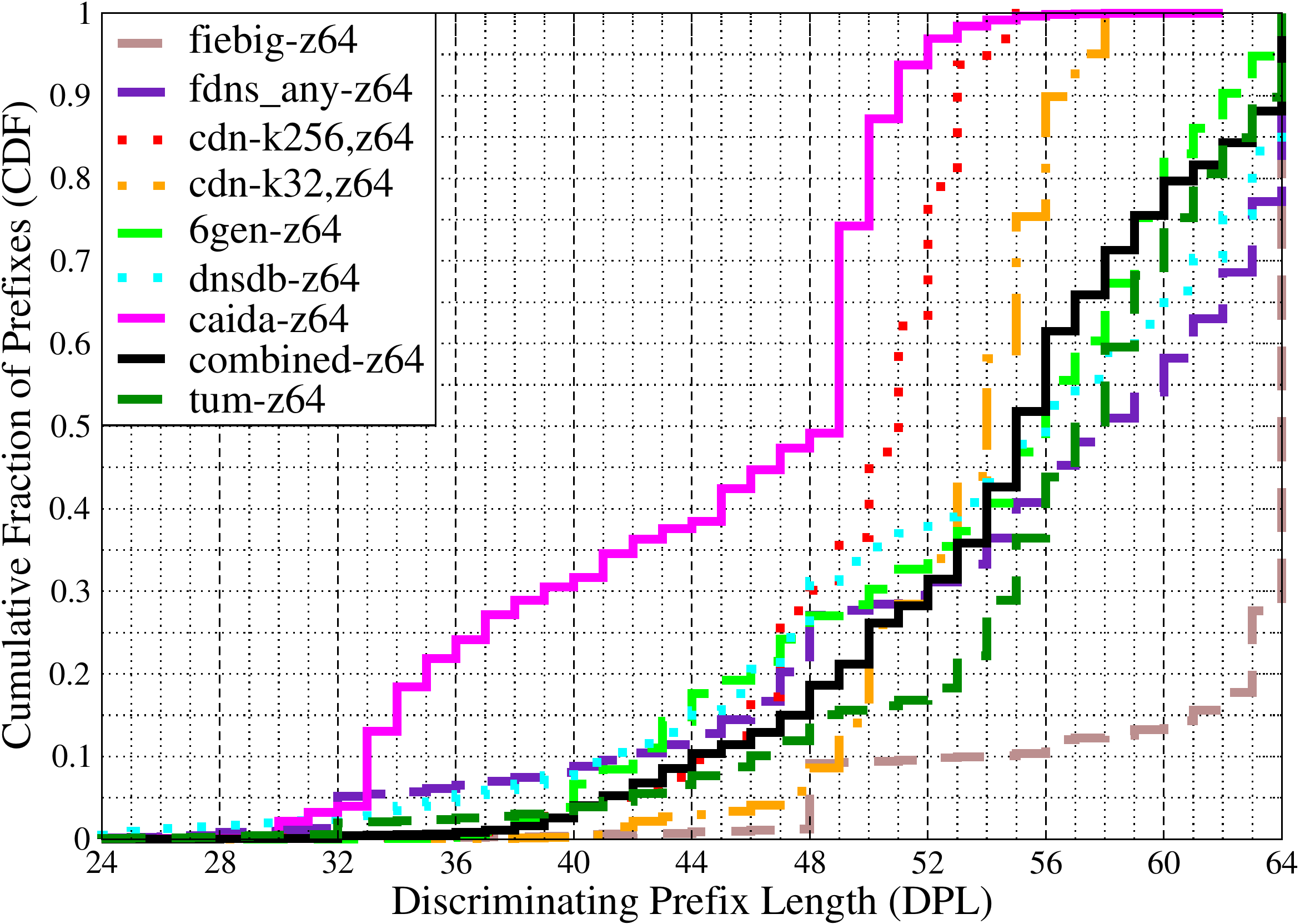}
  \caption{DPL distribution for addresses in each set}
  \label{fig:targDPL}
  \end{subfigure}
  \begin{subfigure}{0.42\textwidth}
  \centering
  \includegraphics[width=\linewidth]{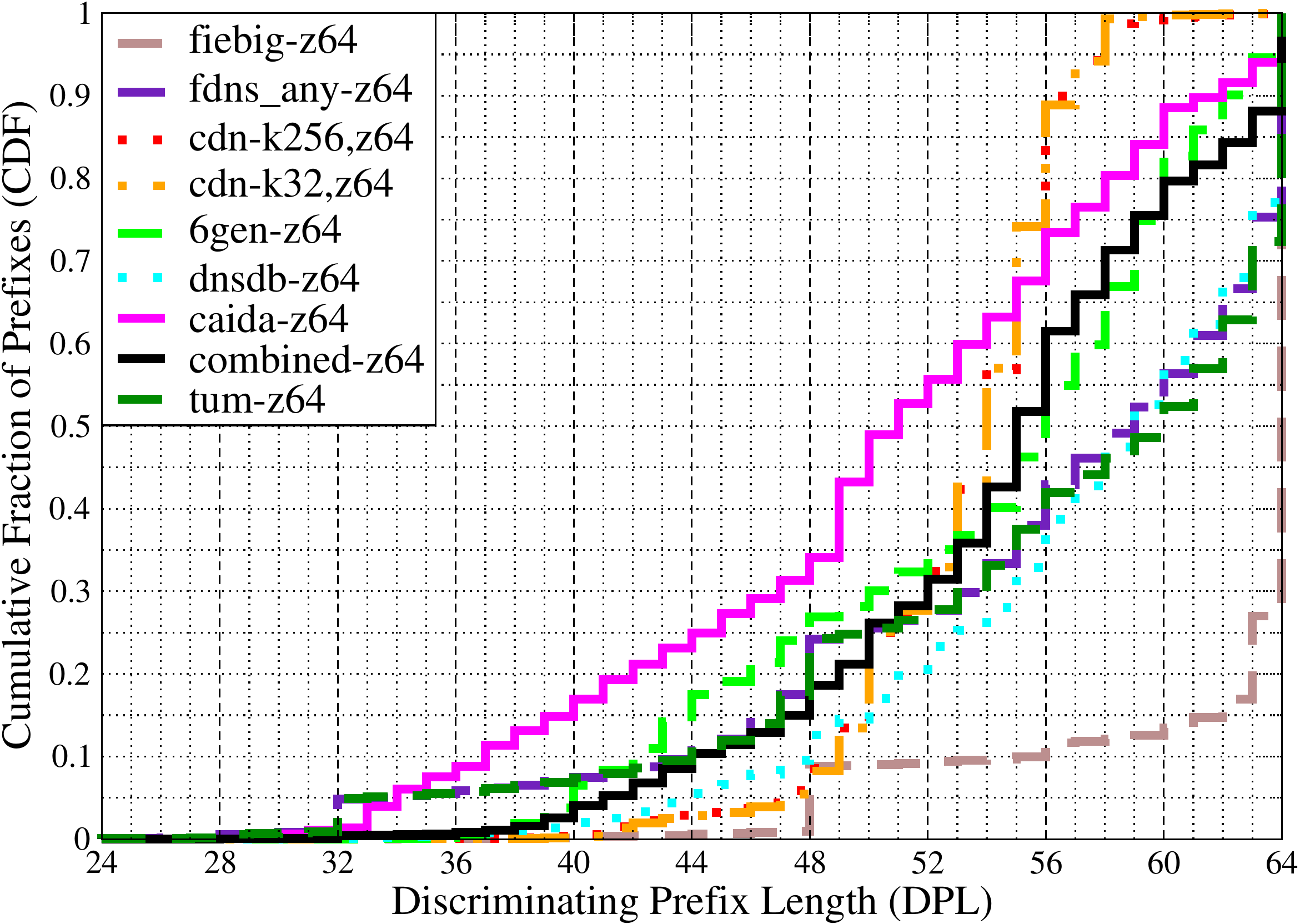}
  \caption{DPL distribution when sets are considered together}
  \label{fig:targCombinedDPL}
  \end{subfigure}
  \caption{Discriminating Prefix Length (DPL) Distributions for target sets (CDF)}
  \label{fig:targDiscriminatingPower}
  \vspace{-4mm}
\end{figure}

\input{targetListProperties_table_scaled}

\subsection{Target Set Characterization}
After \replaced{generating}{aggregating the seeds into} our target sets
\added{as in Section~\ref{sec:tas} and Figure~\ref{fig:targetselect}}, 
we characterize and compare them across multiple dimensions in 
Table~\ref{tab:targets}.  Unique targets refers to the number of targets in the set 
after duplicates have been removed.  Exclusive (Excl) features are those found only in 
one of the sets and not any other set.  We find that most of the sets have some
fraction of targets that are not found in the public BGP IPv6 routing tables, and in
some cases (\eg Fiebig) this fraction is significant, so the ``Routed Targets'' and ``Exclusive Routed Targs'' 
columns characterize only the subset of targets that appear in BGP.  Additionally we note
that some target sets, despite a large number of targets, are concentrated in only a few
BGP prefixes and ASNs, based on
the number of BGP prefixes and ASN represented in each set.

As mentioned in Section~\ref{sec:tas}, some seed lists, and consequently the corresponding 
target sets, are not independent of others.  When computing features that are exclusive to any of the 
first 14 sets listed \added{in Table~\ref{tab:targets}} (CAIDA through 6Gen), we do not consider the  Combined \deleted{sets we created, }or \deleted{the }TUM sets, since they
would \replaced{mask}{cancel out} the exclusive contributions of their respective subsets.  For the TUM sets, however,
we do show the features that are found in that set exclusively.  The \replaced{Combined and Total}{``Total''} sets \added{appear here} merely to \deleted{allow us to }\replaced{show}{refer to}
the total number of unique features that exist across all \deleted{the }sets; \replaced{they}{those sets} are not independently probed\added{ as their own campaigns}.

Figure~\ref{fig:targ-stack} shows the \deleted[id=2]{break-out of }features contributed exclusively by each z64 target set.
Clearly there are a few large players in terms of number of targets/routed targets, however 
this does not strongly correlate to representation in BGP \added{prefixes (Pfx)} or ASNs.  Since the vast majority
of \deleted{BGP }prefixes and ASNs are represented in more than one target set, we provide an
alternate inset view of those two features with the shared contribution removed.  From this we
can \deleted{clearly }see the lack of correlation between target set size and those \replaced{BGP}{two} features.

\subsubsection{Discriminating Power}
\label{sec:DPL}

In this work we rely heavily on the notion of address' discriminating
prefix length (DPL).\footnote{Kohler \etal\cite{DBLP:conf/imc/KohlerLPS02} introduced the term
``discriminating prefix length.'' It has been employed in the
structure analysis of both active and passive measurements.}
An address' DPL is the first (leftmost) bit at which it differs
from \replaced{its}{it's} nearest address accompanying it in a (sorted) set, \eg one
of our target sets. From left to right (high to low order) bits across
the address, the DPL is how far one must compare addresses, bit by
bit, to discriminate
them from each other, for instance how a primitive router might
determine which way to forward traffic if two addresses required
different treatment.  As such, the DPLs of addresses in a set
capture their proximity to each other -- the higher the DPLs
the closer the addresses are -- and, when two addresses are
topologically heterogeneous, \eg in different subnets, the addresses'
DPL is a lower bound on their respective subnets' prefix length.
We will first employ DPL to characterize our target sets. Later,
in Section~\ref{sec:subnetdiscovery}, we \deleted{will }use DPL when discovering
network topology from trace results.

Figure~\ref{fig:targDPL} explores the potential power of each target
set to discriminate addresses and subnets, on its own. Note how
the distribution of discriminating prefix lengths characterizes
a target set. For instance, about 50\% of the {\tt caida-z64}
target addresses have a DPL less than 48, \ie they do not
share the same top 48 bits, and therefore are not covered by the same /48 prefix.
In contrast, over 70\%
of the {\tt fiebig-z64} target addresses have DPL of 64, meaning
the addresses share the top 63 bits, \ie are very near one another.

Figure~\ref{fig:targCombinedDPL} explores the increased potential
discriminating power that each target set brings when they are used
in combination. For example, the {\tt caida-z64} distribution
shifts right, meaning that some addresses from other target
sets are interleaved amongst \deleted{the} its target addresses, thus
yielding more power to discover router hops on paths to
more (specific) routed prefixes or subnets. One might say addresses
in some target sets cleave apart addresses in others, yielding a
more powerful combined target set in depth as well as breadth.
In contrast, note that the distribution of {\tt fiebig-z64} is
unaffected by the combination. This is largely because it has
densely-arranged target addresses, 89\% of which are 
unique (see Table~\ref{tab:targets}) and are not interleaved
with those of other target sets.

Note Figure~\ref{fig:targCombinedDPL} allows us to make {\em
predictions} about the discriminating power of the target sets in
combination with each other as measured by a shift rightward to
higher DPL values.  For instance, {\tt cdn-k256,z64} when combined
with {\tt cdn-k32,z64} discriminating power shifts to that of
latter, which happens to contain 94\% of the targets in the former.
Also, combining {\tt fdns\_any-z64} with {\tt tum-z64} converge to similar
discriminating power. Presumably this is a side-effect of the latter
largely including the former. Indeed, 88\% of the targets in {\tt
fdns\_any-z64} are contained in {\tt tum-z64}.  Additionally, {\tt dnsdb-z64}
roughly converges with {\tt tum-z64} and {\tt fdns\_any-z64}\deleted{, and given}. Given the latter \deleted{that
it }is also DNS-based, it is plausible \deleted{to conclude }that it explores \added{some} \replaced{some common address space}{similar IP-space}.

Also noteworthy is what does not change between Figures~\ref{fig:targDPL} and~\ref{fig:targCombinedDPL}.  None of the ``large'' target sets ({\tt cdn-k32,z64}; 
{\tt 6gen-z64}; or {\tt tum-z64}) shift noticeably to the right.  From this we draw two 
insights.  First, combining significantly smaller sets with large sets has no significant
impact on the potential discriminating power of the large set (unsurprisingly), even 
when significant interleaving occurs.  Second, none of our large sets interleave significantly
with each other.  This is both good and bad, meaning that they are complementary in terms of 
the regions of \vsix space explored, with the tradeoff being that they do not reinforce 
one another to enable addition depth of topology discrimination.  If two similarly sized
\emph{and} interleaved target sets were combined, we would expect the combined potential 
discriminating power to be higher than either set individually.

To reiterate, at this stage we are only discussing \emph{potential} 
discriminating power because these are only target sets; they are 
not empirically discovered.  While some target sets \replaced{capture prefixes}{are based on addresses} of real hosts,
\deleted{but }others are synthetically generated.


%% file: targetListProperties_table_scaled.tex

\begin{table*}[t]
  \centering
  \small
  \caption{Target Set Properties}\label{tab:targets}
  \vspace{-3mm}
  \begin{tabular}{|c|c||r|r|r|r|r|r|r|r|r|}
    \hline
    \multirow{2}{*}{\textbf{Name}} & \multirow{2}{*}{\textbf{Agg}} & \multicolumn{1}{c|}{\textbf{Unique}} & \multicolumn{1}{c|}{\textbf{Exclusive}} & \multicolumn{1}{c|}{\textbf{Routed}} & \multicolumn{1}{c|}{\textbf{Exclusive}} & \multicolumn{1}{c|}{\textbf{BGP}} & \multicolumn{1}{c|}{\textbf{Exclusive}} & \multicolumn{1}{c|}{\multirow{2}{*}{\textbf{ASNs}}} & \multicolumn{1}{c|}{\textbf{Exclusive}} & \multicolumn{1}{c|}{\multirow{2}{*}{\textbf{6to4}}} \\
    &  & \multicolumn{1}{c|}{\textbf{Targets}} & \multicolumn{1}{c|}{\textbf{Targets}} & \multicolumn{1}{c|}{\textbf{Targets}} & \multicolumn{1}{c|}{\textbf{Routed Targs}} & \multicolumn{1}{c|}{\textbf{Prefixes}} & \multicolumn{1}{c|}{\textbf{BGP}} &  & \multicolumn{1}{c|}{\textbf{ASNs}} & \\
    \hline\hline
    \multirow{2}{*}{\textbf{CAIDA}} & z48 & 78.3k & 26.4k & 76.2k & 25.4k & 47.5k & 1.4k & 14.4k & {254} & {0} \\
    \cline{2-11}
    & z64 & 105.2k & 56.6k & 102.7k & 55.2k & 47.6k & 1.4k & 14.4k & {254} & {0} \\
    \hline
    \multirow{2}{*}{\textbf{DNSDB}} & z48 & 93.8k & 9.3k & 93.3k & 9.4k & 36.5k & {360} & 12.8k & {73} & {80} \\
    \cline{2-11}
    & z64 & 233.0k & 101.1k & 231.9k & 101.0k & 36.7k & {442} & 12.8k & {72} & {80} \\
    \hline
    \multirow{2}{*}{\textbf{Fiebig}} & z48 & 102.2k & 85.4k & 45.1k & 28.8k & 6.0k & {8} & 3.9k & {2} & {0} \\
    \cline{2-11}
    & z64 & 1.0M & 898.7k & 576.9k & 468.8k & 6.0k & {11} & 3.9k & {2} & {0} \\
    \hline
    \multirow{2}{*}{\textbf{FDNS}} & z48 & 228.7k & 171.8k & 193.0k & 136.7k & 13.4k & {17} & 7.7k & {5} & 88.4k \\
    \cline{2-11}
    & z64 & 746.9k & 566.3k & 709.9k & 530.4k & 13.5k & {34} & 7.7k & {6} & 88.5k \\
    \hline
    \multirow{2}{*}{\textbf{CDN-k256}} & z48 & 162.4k & 2.6k & 162.3k & 2.6k & 3.3k & {0} & {589} & {0} & {160} \\
    \cline{2-11}
    & z64 & 396.7k & 19.6k & 396.6k & 19.5k & 3.3k & {0} & {589} & {0} & {160} \\
    \hline
    \multirow{2}{*}{\textbf{CDN-k32}} & z48 & 524.2k & 341.8k & 523.9k & 341.6k & 4.9k & {3} & 1.2k & {0} & 1.4k \\
    \cline{2-11}
    & z64 & 3.2M & 2.8M & 3.2M & 2.8M & 4.9k & {4} & 1.2k & {0} & 1.4k \\
    \hline
    \multirow{2}{*}{\textbf{6Gen}} & z48 & 1.4M & 1.4M & 1.3M & 1.3M & 44.3k & {171} & 13.8k & {17} & {0} \\
    \cline{2-11}
    & z64 & 4.5M & 4.4M & 4.3M & 4.2M & 44.5k & {287} & 13.8k & {18} & {0} \\
    \hline
    \hline
    \multirow{2}{*}{\textbf{Combined}} & z48 & 2.3M & N/A & 2.1M & N/A & 48.3k & N/A & 14.5k & N/A & 89.9k \\
    \cline{2-11}
    & z64 & 9.5M & N/A & 8.8M & N/A & 48.6k & N/A & 14.5k & N/A & 90.0k \\
    \hline
    \hline
    \multirow{2}{*}{\textbf{TUM}} & z48 & 362.7k & 108.4k & 305.3k & 86.0k & 25.6k & {36} & 10.9k & {10} & 91.0k \\
    \cline{2-11}
    & z64 & 2.1M & 1.3M & 2.0M & 1.3M & 25.9k & {158} & 10.9k & {20} & 91.2k \\
    \hline
    \hline
    \multirow{3}{*}{\textbf{Total}} & z48 & 2.4M & N/A & 2.2M & N/A & 48.3k & N/A & 14.5k & N/A & 100.7k \\
    \cline{2-11}
    & z64 & 10.8M & N/A & 10.1M & N/A & 48.8k & N/A & 14.5k & N/A & 100.9k \\
    \cline{2-11}
    & both & 12.4M & N/A & 11.5M & N/A & 48.8k & N/A & 14.5k & N/A & 114.9k \\
    \hline
  \end{tabular}
  \vspace{-3mm}
\end{table*}

%% file: probing.tex

\section{Probing}
\label{sec:method:probing}

We start with Yelling at Random Routers Progressively (\yrp), a randomized high-speed \vfour topology
prober~\cite{imc16yarrp}.  \deleted{In contrast to traditional traceroute
techniques intended to probe individual paths, 
\yrp is specifically designed for large-scale
Internet-wide topology mapping.  
\yrp
spreads topology probes across the network rather than probing the path
to individual destinations sequentially.  It does this by
randomly permuting the space of destination targets and TTLs,
thereby attempting to avoid 
overloading any single router or path.
}
Rather than maintaining large amounts of state, as \deleted{such} randomization
imposes on a traditional traceroute-based prober, \yrp encodes details of the probe (\eg
originating TTL and timestamp) \emph{within} the packet so that
state can be reconstructed from the ICMP reply.  This
encoding thus allows \yrp to be stateless.  \yrp thus
decouples probing from topology construction; the complete set of
responses to any single destination are not in any order and are
intermixed with the responses from all other destinations.  
Based on promising
results measuring \vfour~\cite{imc16yarrp}, we explore adapting these techniques to the \vsix domain
\added{which has both a much larger and sparsely populated address
space, as well as mandated \icmpsix rate limiting}.

\subsection{\yrpsix}


{\bf State encoding.}
IPv6 requires several changes in order to retain the stateless nature
of \yrp. First, \vsix headers have removed some of the fields used by
\yrp in \vfour to encode state, \eg IPID.
While there is less room for encoding within the \vsix
packet header, conversely, \icmpsix affords the advantage of complete
packet quotations.  Rather than having to encode state into the
probe's packet headers such that a partial (\added{28 bytes,} 28B) packet quotation
contains \yrp state, the \icmpsix specification requires as much of the
packet that induced the ``Time Exceeded'' message to be returned as
possible~\cite{rfc4443}.  This allows us to encode and
recover more state (than with IPv4) and ensures that header values remain constant
so all probes for the same destination follow the same
path when load balancing is present~\cite{augustin2006avoiding}.
Further, placing state in the payload removes the need to encode 
data within the transport protocol (for example, \vfour \yrp uses the
TCP sequence number to encode the timestamp), and thus easily
facilitates using multiple transport protocols (\yrpsix supports TCP,
UDP, and \icmpsix).


\begin{figure}[t]
 \centering
 \resizebox{0.95\columnwidth}{!}{\includegraphics{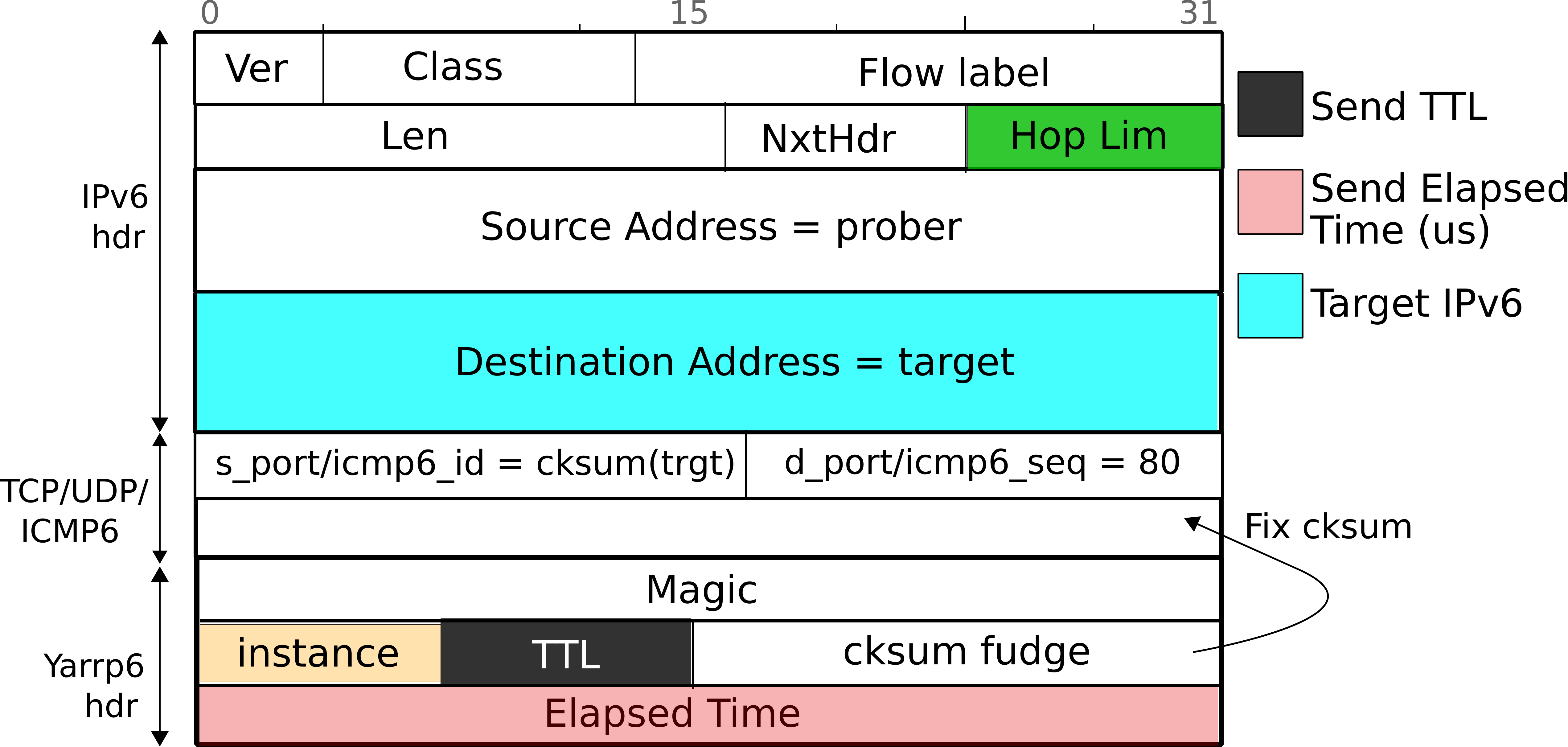}}
 \vspace{-3mm}
 \caption{\yrpsix state: a 12B application payload carries state and
 corrects the checksum such that the \vsix and transport headers
 remain per-target constant.}
 \label{fig:fieldsipv6}
 \vspace{-4mm}
\end{figure}

Figure~\ref{fig:fieldsipv6} depicts \yrpsix state encoding within
the probes it sends.  After the \replaced[id=2]{\vsix}{\vfour} and transport headers is the
\yrpsix payload of 12B.  The \yrpsix payload consists of a 4B magic
number and 1B instance ID to ensure that received \icmpsix packets are
indeed responses to \yrpsix probes, and for the running instance.  A single byte encodes the originating
TTL (``hop limit'').  Four
bytes encode the probe's timestamp to permit round-trip-time
(RTT) computation.

\added{Similar to \vfour paths, Almeida \etal find load
balancing prevalent on \vsix paths they
sample~\cite{almeida2017characterization}.}
We therefore wish to ensure that the packet headers remain constant and
accommodate load balanced paths, however the \added{transport} checksum will be different
for each probe as the TTL and timestamp in the \yrpsix payload
changes.  \added{While the transport checksum is not typically
used for load-balancing TCP and UDP flows, the \icmpsix checksum
is \replaced[id=2]{employed in}{part of the} per-flow load balancing.} We therefore include 2B of ``fudge'' within the \yrpsix
payload in order to ensure that the \added{transport} checksum also remains constant.
Finally, we compute a 2B Internet checksum over the \vsix target address
and, depending on the transport protocol selected, use it for the
TCP or UDP source port or \icmpsix identifier.  This checksum 
\replaced[id=2]{permits detection of modifications to the \vsix target
address, \eg due to middleboxes}{ensures that the quoted \vsix target address is
unmodified}.

{\bf Fill Mode.} 
A consequence of stateless operation
is that \yrp cannot stop probing when it reaches its destination, or
encounters several unresponsive hops in a row (the so-called
``gap-limit'')~\cite{imc16yarrp}.  Instead, the user must select the
probe TTL range
a priori, potentially missing hops if the maximum TTL is smaller than
the path length, or wasting probes if the maximum TTL is too large.
To \deleted{better }accommodate this tension, we add to \yrpsix
a ``fill mode.'' 

Let the user-selected\added{, {\em initial}} maximum probing TTL be $m$.  In fill mode, if
\yrpsix receives a response for a probe sent with hop
limit $h$ where $h \ge m$, it immediately sends a new probe toward the
destination with a hop limit of $h+1$.  While these additional probes
are not randomized, fills are uncommon 
and occur at the tail of the path where the effect of
sequential probing has the least impact.  We explore tuning of \yrpsix
parameters, including fill mode, next.


\begin{figure*}[t]
\begin{centering}
  \begin{subfigure}{0.48\textwidth}
    \centering
    \resizebox{3.4in}{2.0in}{\includegraphics{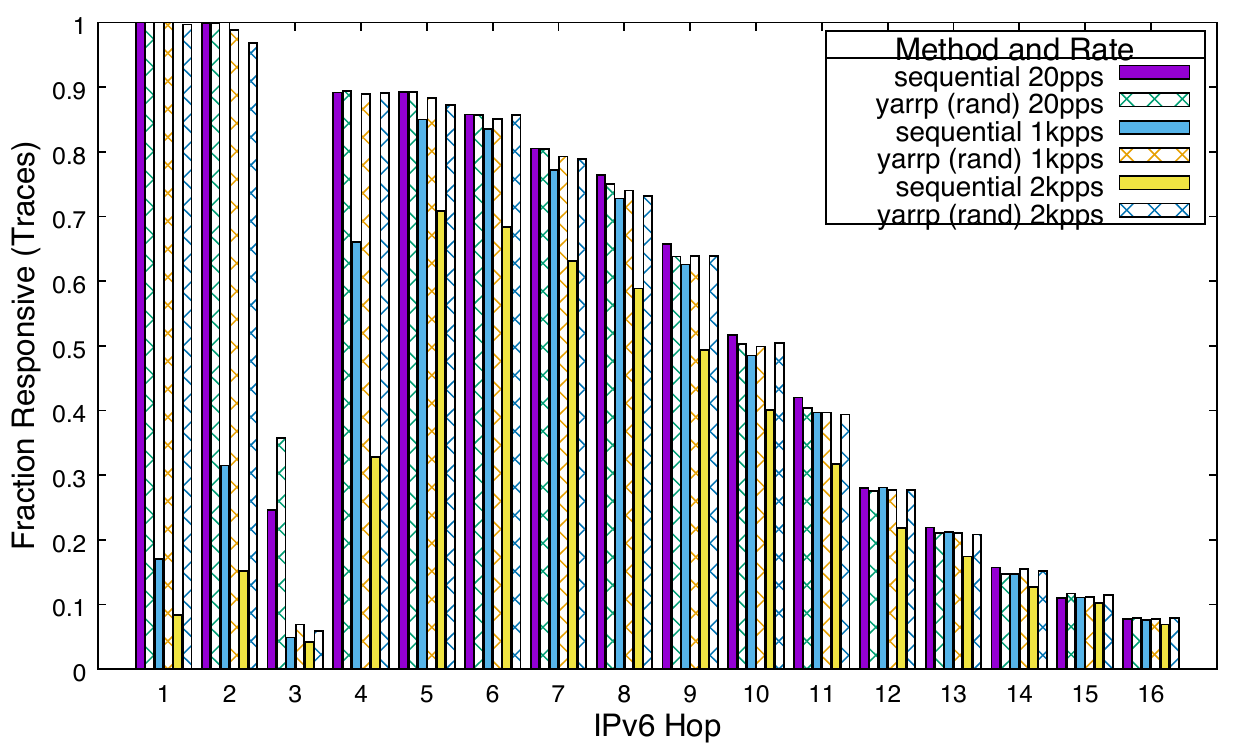}}
    \caption{Vantage: US-EDU-3}
    \label{fig:speed:nps}
  \end{subfigure}
  ~
  \begin{subfigure}{0.48\textwidth}
    \centering
    \resizebox{3.4in}{2.0in}{\includegraphics{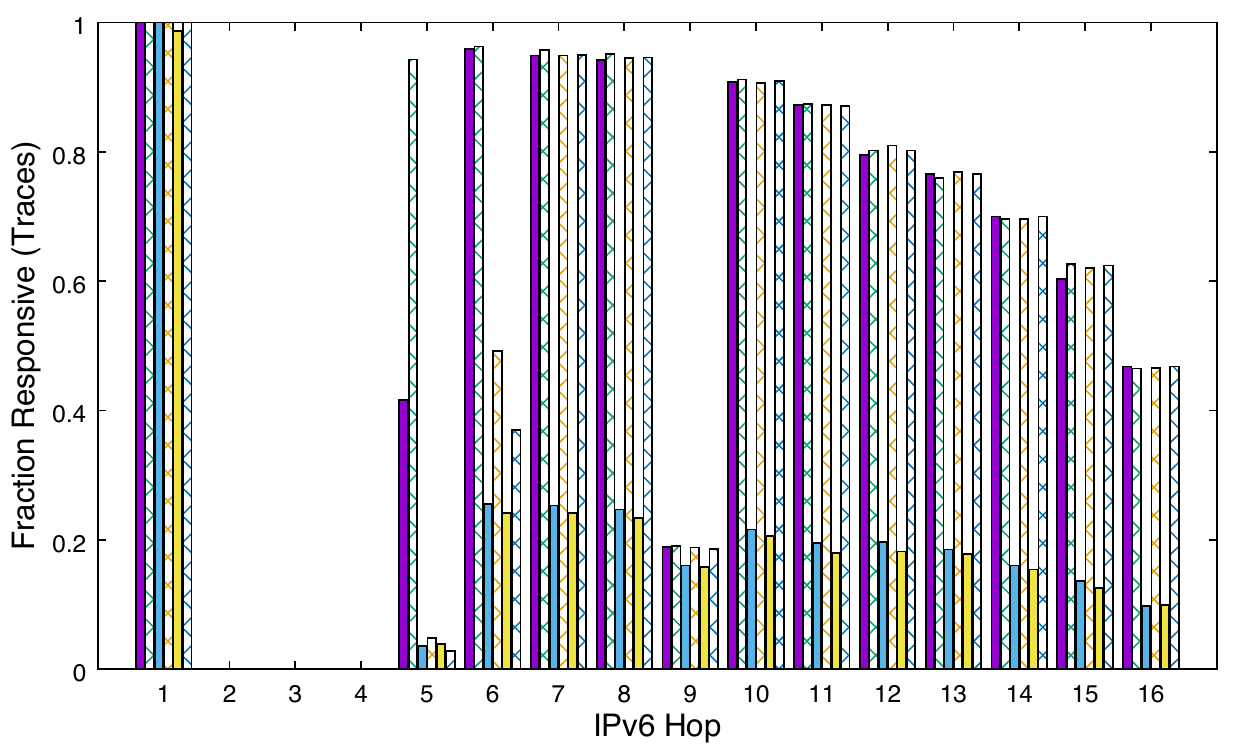}}
    \caption{Vantage: US-EDU-2}
    \label{fig:speed:uo}
  \end{subfigure}
  \vspace{-3mm}
  \caption{\added{Trial Results:} Relationship between probing strategy, rate, and per-hop
           responsiveness at two vantage points.  While \icmpsix
           rate-limiting is clearly evident, randomly permuting the
           probing order universally improves responsiveness.}
  \label{fig:speed}
\end{centering}
\vspace{-4mm}
\end{figure*}

\subsection{Tuning}
\label{sec:probing:tuning}

In addition to selecting vantage points and targets, active topology
discovery requires choosing the probing protocol, maximum TTL, speed,
and other parameters. 
This section explores these parameters 
to better understand
the tradeoffs and to guide our probing strategy.

\added{An evaluation metric we employ is the
number of discovered ``interface addresses.''  Henceforth, we define an interface
address to be a unique \vsix source address of received \icmpsix 
messages, and do not attempt to correlate these with interfaces or
among routers.  In particular, a packet forwarder (\eg router)
has multiple interfaces, including virtual loopbacks, and each
interface can have multiple addresses.  While \cite{rfc4443} specifies
that nodes should choose a unicast \vsix source address for \icmpsix
packets according to the packet's destination, other behaviors may
exist.}

{\bf Protocol.}
Any \vsix packet can be used for active topology probing,
however the well-known prevalence of middleboxes and firewalls
suggests that different \replaced{packet types}{transport protocols}, \eg TCP SYN, TCP ACK,
UDP, or \icmpsix, can yield different results depending on whether these
are blocked along a path or whether an in-path device maintains
connection state.  Luckie \etal studied the influence of transport
protocol for \vfour and found that ICMP-Paris reaches the most 
destinations, while UDP based methods infer the greatest number of
IP links~\cite{luckie08}.  CAIDA's production active
topology discovery \cite{caida-topo, caida-topov6} utilizes ICMP-Paris
for both \vfour and \vsix.

To the best of our knowledge, there is no published equivalent
analysis of the effect of transport protocol on \vsix active topology
discovery.  We therefore mounted \added{trial} probe campaigns from two of our
vantage points on February 1, 2018 using the CAIDA target set.  We
use the same permutation seed and targets to probe using TCP, UDP, and
\icmpsix.  To mitigate the possible effects of any rate-limiting, we
probed at only 20pps. 

On average, we see that probing with \icmpsix results in $\sim$2.2\%
and 2.1\% more discovered interface \added{addresses} than using UDP and TCP
respectively.  Interestingly, \icmpsix probes produce on average
13.6\% and 24.3\% more non-``Time Exceeded'' \icmpsix responses than UDP
and TCP respectively, suggesting that these probes are penetrating
deeper into the network.  Given these observations, we send \icmpsix
probes
for the remaining experiments.

{\bf Speed.}
As discussed previously, probing speed has a potentially greater
impact on discovery in \vsix as compared to \vfour due to mandated
rate-limiting.  Toward understanding the behavior of various
probing techniques and speeds, we mount \added{trial} probing 
campaigns to the CAIDA target set on April 27, 2018 from our vantage
points.  Figure~\ref{fig:speed} shows the \added{per-trace} fraction of responses 
received \replaced[id=2]{as a function of responding \vsix hop for speeds of 20, 1000, and
2000 pps, and using }{Further, we explore}two probing strategies: randomized (as
implemented in \yrpsix) and sequential  (as
implemented in scamper~\cite{luckie10scamper}). 
Scamper running in
\icmpsix Paris mode represents the current state-of-the-art topology
probing tool and technique used in production systems.
Note that because the number
of responsive interface \added{addresses} decreases as the hop distance increases, it
is necessary to compare the relative performance of \yrpsix and
\replaced[id=2]{sequential}{scamper} at different speeds.   

We observe markedly different results 
\replaced[id=2]{between sequential and}
{from using
scamper versus the
permutation of} \yrpsix at speeds above 20pps, especially nearer to the
vantage point.  For instance, in Figure~\ref{fig:speed:nps}, while 
\replaced[id=2]{sequential}{scamper} and \yrpsix have nearly identical response rates at 20pps,
\yrpsix yields a 100\% response rate from the first hop for 1000 and
2000pps as compared to less than 20\% and 10\% for
\replaced[id=2]{sequential}{scamper}.  Across
both vantage points, we observe better performance with \yrpsix at
all hops than achieved with \replaced[id=2]{sequential}{scamper} at higher probing rates.

\added[id=2]{Investigation of timing from packet captures of the two tools 
shows per-TTL bursty behavior in the sequential prober implementation,
a behavior that persists as traces remain synchronized.  In addition
to \yrp's randomization strategy, the rate-limiting effects could 
also possibly be mitigated with a different transmission behavior, an
observation recently made in~\cite{MAPRGRateLimitingIPv6}.}

Also of note are the variety of rate-limiting behaviors implemented by
routers.  For instance, hop 3 of Figure~\ref{fig:speed:nps} and hops 5
and 9 of Figure~\ref{fig:speed:uo} appear to implement more aggressive
rate limiting as compared to the other hops.
Further, while not pictured, we find one hop near one of our vantage
points that only responds with time exceeded messages when \icmpsix is
used as the probe type.  Because subsequent hops respond, we
conjecture that this behavior is due to some form of state maintenance
for security reasons.

\added{
{\bf Probe order.}  
While sequentially increasing TTL tracing clearly
induces \icmpsix rate limiting, there are a variety of potential probe
order strategies that can help mitigate the effects of rate limiting,
including \yrpsix's randomization.  Since we first observed the
effects of \icmpsix rate limiting on active topology
scans~\cite{aims17yarrp6}, other researchers have recently reported
similar findings and explored different probing
strategies~\cite{MAPRGRateLimitingIPv6}.
}

\added{
The tree-like structure of the network implies that hops closest to
the vantage will experience the most probing traffic, and it follows
that we observe these nearby hops exhibiting the most rate-limiting.
Thus, a natural strategy to explore is
Doubletree~\cite{donnet2005efficient} which first observed that paths
exhibit significant redundancy in their initial hops.  Doubletree
chooses an intermediate starting TTL and probes forward (increasing
TTL) and backward (decreasing TTL) until it receives a response from
an interface it has previously observed.  Doubletree then fills in
missing portions of the path based on previous results.
}

\added{
To better understand Doubletree's performance relative to \icmpsix
rate limiting, \added{as a trial,} we also probe the CAIDA target set from our vantage
points at various packet rates using scamper's Doubletree
implementation.  While Doubletree induces less rate limiting than
traditional traceroute methods, we observe an unexpected effect: when
rate limiting causes a hop to be non-responsive, Doubletree continues
probing backward\added[id=2]{ in the TTL space}.  Thus, as the token buckets on the initial hops
drain, Doubletree continues to probe them, causing them to remain
empty.  Notwithstanding this behavior, Doubletree has two fundamental
limitations. First, it is necessary to select the intermediate
starting TTL, a parameter that must be heuristically estimated and set
for each vantage point.  Second, using results from previous traces to
fill in hops that were not probed can lead to erroneous
path\replaced[id=2]{
inference, either due to violations of destination based routing}{s}, as
described in~\cite{Flach:2012:QVD:2398776.2398804} \added[id=2]{or path
changes}.  We therefore
believe that \yrpsix provides a compromise between completeness
and scalability that is well-suited to Internet-wide \vsix topology 
studies.
}

\added[id=2]{Finally, we note that \yrp also includes the ability to employ
a similar heuristic as Doubletree where it can maintain state over its
local responsive neighborhood (\eg router interface hops below a
configurable TTL).  In this mode, \yrp maintains a per-TTL timestamp
of the last time a probe was sent and the last time a probe elicited a
new interface address.  If no new
addresses are discovered within a
configurable window of time, \yrp skips future probes for that TTL.
While this work purposefully takes an exhaustive stateless approach,
in future research we plan to experiment with \yrpsix's neighborhood
enhancement.}

{\bf TTL range.}
As described previously, \yrpsix's fill mode
balances the choice of a maximum probe TTL with
discovery rate and volume of probing.  Recall that the maximum TTL
must be selected in advance as part of the permutation process.  Thus,
a large maximum TTL will potentially waste probes, while a small
maximum TTL will potentially miss hops.  Fill mode allows us to select
a lower maximum TTL, thereby lowering probing volume, while missing
fewer hops.

To quantify this tradeoff, we explore the use of different \yrpsix 
maximum TTL values \replaced{in a final set of trials}{when} probing the CAIDA target set on May 2, 2018. 
In \replaced{these trials}{all experiments}, fill mode continues probing past the maximum
TTL so long as responses are received, up to a maximum hop limit of
32.  Table~\ref{tab:fillmode} shows the number of probes, probes
resulting from fills, unique interface \added{addresses} discovered, and the interface
\added{address} yield (\added{addresses} discovered per probe).  Note that a single
non-responsive hop past the maximum TTL will cause fill mode to stop.
Thus, we observe that the number of fills for a maximum TTL of four is 
much less than for a maximum TTL of eight simply because hop five did
not respond.
Using a maximum TTL of 16 produces the highest yield; we therefore 
use this for \replaced{all subsequent campaigns (\ie results in Section~\ref{sec:results})}{the remainder of the experiments}
in order to achieve the most efficient use of probing.

\input{fillmode_table_scaled}

\subsection{Ethical Considerations}

Performing large-scale, Internet-wide active
measurements requires careful consideration of the experimental
methodology to avoid causing harm.  First and foremost, we obtained explicit permission from
the networks hosting the vantage points in our study.  Second, we opt
to send \icmpsix probes as they are relatively innocuous compared to UDP
and TCP, are used by the existing Ark and RIPE platforms thereby
facilitating direct comparison, and yield the most responses as
discussed in Section~\ref{sec:probing:tuning}.  Third, although capable of much higher
rates, we run \yrpsix at 1kpps both to minimize rate-limiting
(Section~\ref{sec:probing:tuning}) and to maintain low network load. (Note that
\yrp's randomization naturally spreads load.)
Fourth, because our goal is to discover \vsix router addresses, not
end hosts, we use the fixed pseudo-random IID for all campaigns,
which is unlikely to be that of an active \vsix host.  We show in
Section~\ref{sec:method:xfrm} that using this IID has 
\replaced{negligible}{neglibible} effect on topology discovery as compared to 
the {\tt ::1} IID.

Finally, we follow best
practices for good Internet citizenship by making an informative web
page, along with opt-out instructions, available at the source address
of our probes~\cite{durumeric2013zmap}.  Over the course of our
probing campaigns, we received two opt-out requests with which we
immediately complied.


%

%% file: fillmode_table_scaled.tex

\begin{table}
 \small
 \caption{Fill Mode \added{Trial Results}}\label{tab:fillmode}
 \vspace{-3mm}
  \centering
  \begin{tabular}{|c|r|r|r|r|}\hline
    \textbf{MaxTTL} & \multicolumn{1}{c|}{\textbf{Probes}} & \multicolumn{1}{c|}{\textbf{Fills}} &\multicolumn{1}{c|}{\textbf{Int Addrs}} & \multicolumn{1}{c|}{\textbf{Yield \%}} \\
    \hline\hline
    4  & 375.6k & 96.4k & 271  & {0.1} \\\hline
    8  & 751.2k & 213.5k & 11.3k & {1.2} \\\hline
    16  & 1.5M & 251.5k & 39.1k & {2.2} \\\hline
    32  & 3.0M & 0  & 54.1k & {1.8} \\\hline
  \end{tabular}
  \vspace{-3mm}
\end{table}

%% file: results.tex

\section{Results}\label{sec:results}

\input{results_table_scaled}

We present the following results:
\pt{i} an analysis of the power of the target sets to yield
interface addresses in \yrpsix campaigns;
\pt{ii} comparisons of \yrpsix results to other probers performing
similar and different trace campaigns;
\pt{iii} detailed results of high-frequency \yrpsix campaigns
launched from three vantage machines each with 18 different
target sets (54 in total) on May 14, 2018 and summarized in
Figure~\ref{fig:results_stacked} and Table~\ref{tab:results}.
The top line in the table is the combined result across all
\replaced{three}{(3)} vantages and all campaigns per vantage.

\vspace{-1mm}
\subsection{Topology}

In terms of overall discovery, the two best performing target sets 
are {\tt cdn-k32} and {\tt tum}.  Not only do they produce the largest absolute
numbers of interfaces, they continue to reveal new addresses
throughout the entire probing duration.  As shown in 
Table~\ref{tab:results}, they are largely complementary and 
contribute the two largest shares of interfaces exclusively discovered
by single target sets.

\begin{figure}[h!]
 \centering
 \includegraphics[width=\linewidth]{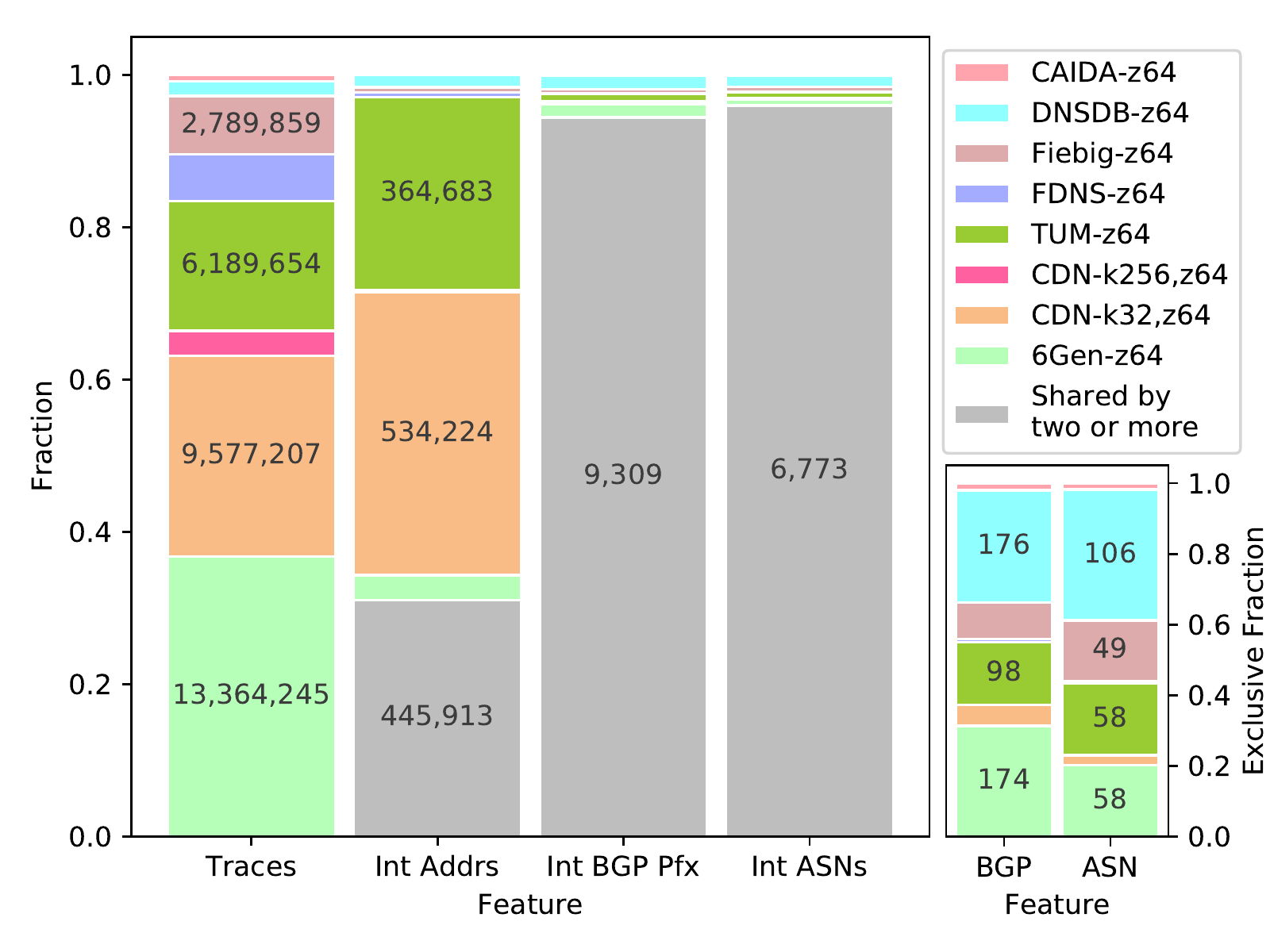}
 \vspace{-6mm}
 \caption{Selected Result Features of \yrp Campaigns (corresponding to Table~\ref{tab:results}).}
 \label{fig:results_stacked}
 \vspace{-3mm}
\end{figure}

Figure~\ref{fig:results_stacked} lets us compare the fraction of total traces 
performed for each target set \deleted{, with}\replaced{by}{with} features of the router addresses
discovered as a result of those traces.  In the small axes on the right
of the figure, we isolate just the fraction of exclusive BGP prefixes, and exclusive ASNs for each set, 
since that is obscured by the shared portion on the main figure.




Rightmost in Table~\ref{tab:results}, as with the seeds in Section~\ref{sec:method},
we used {\tt addr6} to classify the resulting \replaced{interface}{router hop} addresses
discovered across all trace campaigns. Surprisingly, we find
\deleted{very} many EUI-64 \deleted{router hop} addresses, 651.4k or 45\% of all
\replaced{interface}{router hop} addresses. These are labeled ``EUI-64: Int Addrs'' in
Table~\ref{tab:results}, most prominently yielded by the
{\tt tum} (53\% EUI-64 results) and {\tt cdn-k32} (39\% EUI-64 results) campaigns.
\added{The accompanying percentage shows the proportion of EUI-64
\replaced{interface}{router hop} addresses within each campaign.}
Of these EUI-64 router addresses, 59\%
are from one of just two manufacturers; 99.9\% of each of those 
address are in just two ISP networks, each in
different countries.  In both cases, WWW content suggests they are
Customer Premises Equipment (CPE) routers in ostensibly large,
homogeneous \vsix deployments. \added{Offering further evidence, the last column
labeled "EUI-64: Path Offset" explores the distribution hop positions
for EUI-64 \replaced{interface}{hop} addresses, as a negative offset from end of path.
In all CDN campaigns, we see the 5th percentile and median are 0,
meaning 95\% of their EUI-64 addresses are the last hop on path.
For the TUM campaigns, 50\% of their EUI-64 addresses are
the last hop on path (median is 0) while 95\% are one of the last 3 hops
(5th percentile is -3).
This topological difference is presumably the result of increased
heterogeneity of TUM versus CDN targets.}

It is not a coincidence that these two target sets also have
highest overall yield and highest yield of exclusive 
\replaced{interface}{router hop} 
addresses \added{(as seen in the ``Excl Int Addrs'' column)}. This is due, in large part, to elicited \icmpsix 
responses from these CPE routers, one target set yielding one
manufacturer's routers in one ISP's, and the other ISP containing
a different router manufacturer.


\begin{figure}[h!]
 \centering
 \includegraphics[width=\linewidth]{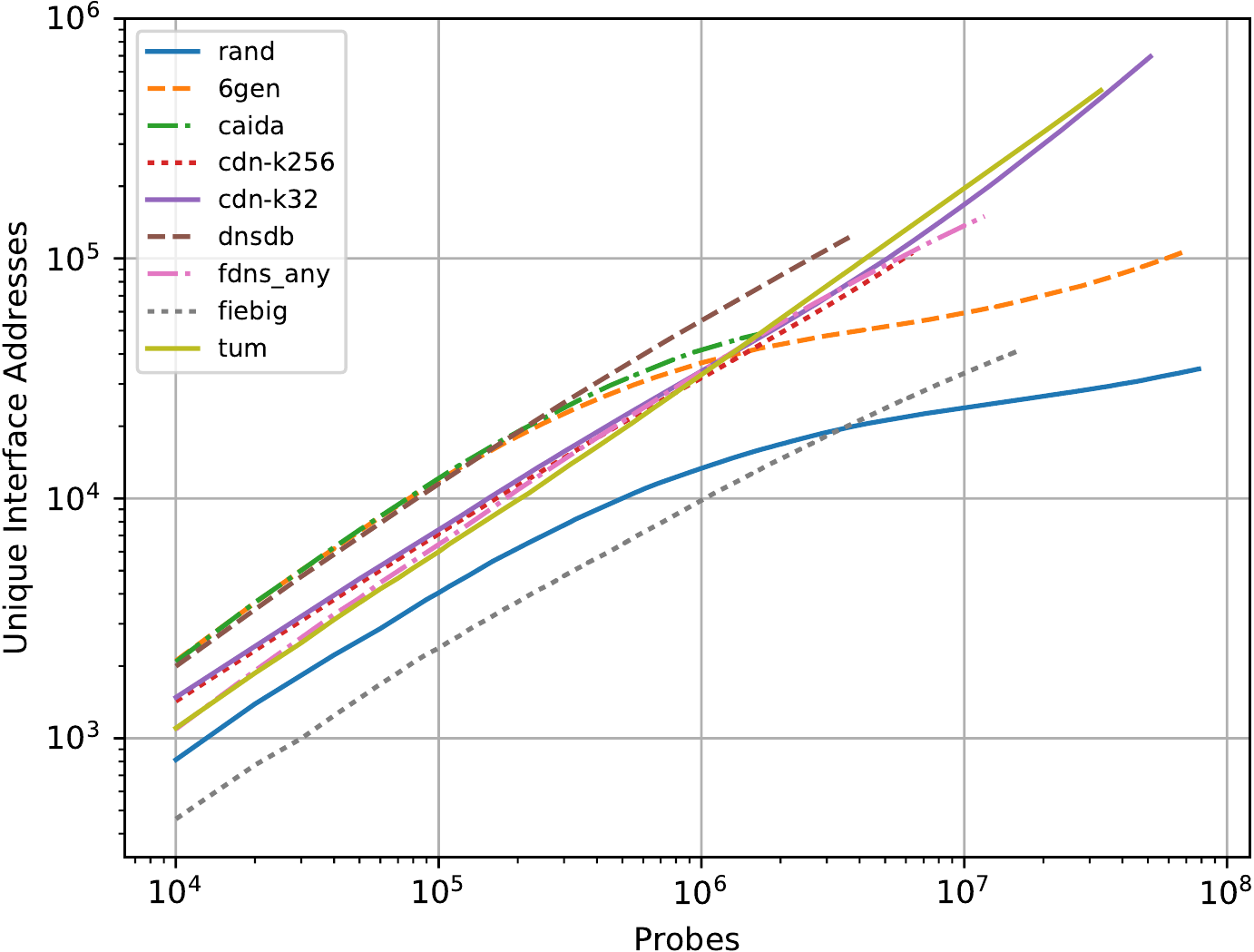}
 \vspace{-6mm}
 \caption{Address discovery power per z64 target set
 vs.\ probe packets emitted (from EU-NET \added{vantage} in Table~\ref{tab:results}).}
 \label{fig:power}
 \vspace{-3mm}
\end{figure}

\vspace{-1mm}
\subsection{Target Sets Power}

Central to our study is evaluating trace target strategies to not only
maximize topological discovery, but also efficiency within an
address space too large to be exhaustively probed.  In
Figure~\ref{fig:power}, we examine the relationship between each z64
target set's count of interface addresses discovered via probing from
the EU-NET vantage point, and the number of probe packets
required.\footnote{Equivalently, at a fixed probing rate, this plot
shows discovery as a function of time.}  \replaced{We}{Note that we} observe
qualitatively similar results from the other vantage points, and
choose to focus on the EU-NET results due to space constraints.

The current state-of-the-art strategy, employed by both CAIDA and RIPE
for their production \vsix mapping, of tracing to the {\tt ::1} address
(plus a random address in CAIDA's case)
within routed \vsix BGP prefixes performs best in the initial stages
of the probing, but suffers a noticeable flattening in discovery past
300k packets (note the log-log plot scale).  CAIDA's discovery peaks
at fewer than 100k interfaces after $\approx$2M probes as it exhausts
the target set.  This dichotomy of performing well initially, but
falling well short of the absolute number of interfaces discovered via
other target sets, illustrates that \replaced{this BGP-based}{CAIDA's} strategy provides breadth,
but lacks the specificity to discover \vsix subnetting where
significant topology exists \added{in depth}.

As a second baseline of a BGP-informed strategy, we probe randomly
generated \vsix addresses that are routable.  Unsurprisingly, this
unguided target selection performs poorly with a precipitous drop in
newly discovered interfaces after $\sim$1M probes.  However,
random outperforms Fiebig prior to this point, largely due to its 
high degree of clustering (evident in the DPL of
Figure~\ref{fig:targDiscriminatingPower}).
Similarly, the 6gen target set provides a high interface yield
at the onset of probing, but flattens past 1M probes.  In fact, the
shape of the 6gen curve closely mirrors random, but with a fixed
positive offset.  

In contrast, the overall discovery rate is higher and linear for
the {\tt tum} and {\tt cdn-k32} synthesized target lists, implying that these
provide the most power.  Although they have similar discovery
yields, {\tt cdn-k32} finds $\sim$200k more addresses from this
vantage.

\subsection{Validation}\label{sec:validation}

As with similar Internet-wide active measurement studies,
\replaced{validation of our method is complicated by limited availability
of ground truth and by limited seed address collections, in part, both
due to privacy concerns}{comprehensive ground truth is not available}.
To place our results in
context, we therefore consider our discovered \vsix topologies
relative to those available from production \vsix active traceroute
systems, namely CAIDA's Ark~\cite{caida-topov6} and RIPE
Atlas~\cite{ripeatlas}.  For both, we gather the complete set of
traceroute results for May 18, 2018 available from each vantage point.
Both Ark and Atlas are global platforms, with vantages in different
regions and networks.  While Ark had 65 \vsix-capable vantages at the
time of writing, Atlas had 4,333.  

Within the 24 hour period, Atlas probed 34.3k unique targets and
discovered 103.8k unique interfaces, while Ark issued traces to 349k
targets which revealed 126.8k interfaces.  Notably, while Ark used
6.9M traces, our methodology discovers $\sim$1.3M interfaces -- an
order of magnitude more -- with only approximately twice the number of
traces.



We further compared our results to those 
of a proprietary prober that regularly performs millions of
traces per day~\added{~\cite{MAPRGRateLimitingIPv6}}. With that prober, the {\tt cdn-k32-z64-fixediid}
target set was traced separately on May 3, 2018, \ie emitting
TTL-limited probes toward active WWW client address space. In
contrast to \yrpsix, this prober is stateful, like {\tt traceroute},
and operates by distributing its workload of traces across a set of
machines, here, from one physical and topological location to make
comparison reasonable. Results show the \replaced{interface}{router hop} address yield
difference was within 1.1\%, and other metrics were similar; it
is plausible that vantage connectivity and system resources are responsible for the difference.

Lastly, in Table~\ref{tab:results}, note there is some variation
in \yrpsix results by vantage, despite launching the same
campaigns. While EU-NET and US-EDU-1 show similar yield, $\sim$1.3M
\replaced{interface}{router hop} addresses, US-EDU-2 has lower yield, $\sim$881k,
despite having performed just as many traces.  Our hypothesis
is that this vantage is unusual in that its on-premise path is longer
(as seen in Figure~\ref{fig:speed:uo} and median path length
in~\ref{tab:results}) and thus may warrant a higher TTL value
(per Section~\ref{sec:probing:tuning}).

%% file: results_table_scaled.tex

\begin{table*}[t]
 \small
  \caption{Results of aggregate \yrp campaigns run from three vantages, reverse sorted by Int Addrs yield, \ie sources of ICMPv6
Time-Exceeded messages. Path Offset of EUI-64 Int Addrs is relative to Path Len, 0  being the last Hop Addr on path.}\label{tab:results}
  \vspace{-3mm}
  \centering
  \begin{tabular}{|c|c||r|r|r|r|r|r|r|r|r|r||r|r|r|}
    \hline
    \multirow{2}{*}{\textbf{\yrpsix}} & \multirow{3}{*}{\textbf{Agg}} & \multicolumn{1}{c|}{\multirow{3}{*}{\textbf{Traces}}} & \multicolumn{1}{c|}{\multirow{2}{*}{\textbf{Target}}} &  \multicolumn{1}{c|}{\textbf{Rtr}} & \multicolumn{1}{c|}{\textbf{Excl}} & \multicolumn{1}{c|}{\textbf{Int}} & \multicolumn{1}{c|}{\textbf{Excl}} & \multicolumn{1}{c|}{\multirow{2}{*}{\textbf{Int}}} & \multicolumn{1}{c|}{\textbf{Excl}} & \multicolumn{1}{c|}{\bf Reach} & \multicolumn{1}{c||}{\textbf{Path Len}} & \multicolumn{2}{c}{\bf EUI-64:} & \multicolumn{1}{c|}{\textbf{Path Offset}} \\
     \multirow{2}{*}{\bf Campaign} &  & & \multicolumn{1}{c|}{\multirow{2}{*}{\textbf{Addrs}}} & \multicolumn{1}{c|}{\textbf{Int}} & \multicolumn{1}{c|}{\textbf{Int}} & \multicolumn{1}{c|}{\textbf{BGP}} & \multicolumn{1}{c|}{\bf Int} & \multicolumn{1}{c|}{\multirow{2}{*}{\bf ASNs}} & \multicolumn{1}{c|}{\bf Int} & \multicolumn{1}{c|}{\bf Target} & \multicolumn{1}{c||}{\bf {95}th Perc.} & \multicolumn{2}{c|}{\bf Int} & \multicolumn{1}{c|}{\bf {5}th Perc.} \\
     &  & & & {\bf Addrs} & \multicolumn{1}{c|}{\bf Addrs} & \multicolumn{1}{c|}{\bf Pfxs} & \multicolumn{1}{c|}{\bf BGP} & & \multicolumn{1}{c|}{\bf ASNs} & \multicolumn{1}{c|}{\bf ASN} & \multicolumn{1}{c||}{\bf (Median)} & \multicolumn{2}{c|}{\bf Addrs} & \multicolumn{1}{c|}{\bf (Median)} \\ \hline\hline

{ALL} & both & 45.8M & 12.6M & 1.4M & 0  & 9.9k & 0  & 7.1k & 0  &  40\%  & {20} (11)  & 651.4k & 45\% & {-7} (0)  \\ \hline
EU-NET & both & 15.0M & 12.2M & 1.3M & 136.0k & 9.5k & 236  & 6.9k & 110  &  44\%  & {17} (8)  & 613.0k & 48\% & {-11} (0)  \\ \hline
US-EDU-1 & both & 15.4M & 12.6M & 1.3M & 84.9k & 9.4k & 75  & 6.8k & 31  &  43\%  & {19} (10)  & 602.7k & 48\% & {-9} (0)  \\ \hline
US-EDU-2 & both & 15.4M & 12.6M & 881.4k & 20.7k & 7.4k & 148  & 5.5k & 76  &  33\%  & {21} (15)  & 540.6k & 61\% & {-2} (0)  \\ \hline
{cdn k32} & z64 & 9.6M & 3.2M & 756.6k & 534.2k & 2.0k & 33  & 1.2k & 8  &  52\%  & {18} (12)  & 297.2k & 39\% & {0} (0)  \\ \hline
{tum} & z64 & 6.2M & 2.1M & 582.4k & 364.7k & 7.9k & 98  & 6.0k & 58  &  63\%  & {19} (12)  & 311.2k & 53\% & {-3} (0)  \\ \hline
{cdn k32} & z48 & 1.6M & 524.2k & 203.7k & 9.6k & 1.8k & 4  & 1.2k & 2  &  70\%  & {19} (12)  & 79.8k & 39\% & {0} (0)  \\ \hline
{fdns} & z64 & 2.2M & 746.9k & 185.2k & 9.4k & 6.2k & 5  & 5.0k & 2  &  48\%  & {19} (10)  & 15.4k & 8\% & {-1} (0)  \\ \hline
{dnsdb} & z64 & 698.6k & 233.0k & 154.0k & 23.2k & 7.8k & 176  & 6.0k & 106  &  55\%  & {20} (13)  & 10.1k & 7\% & {-5} (0)  \\ \hline
{6gen} & z64 & 13.4M & 4.5M & 126.4k & 46.9k & 7.1k & 174  & 5.2k & 58  &  27\%  & {31} (12)  & 24.8k & 20\% & {-18} (0)  \\ \hline
{tum} & z48 & 1.1M & 362.7k & 118.3k & 2.1k & 6.6k & 6  & 5.0k & 2  &  41\%  & {18} (9)  & 11.6k & 10\% & {-3} (0)  \\ \hline
{cdn k256} & z64 & 1.2M & 396.7k & 116.9k & 3.4k & 1.2k & 0  & 639  & 0  &  51\%  & {18} (11)  & 28.5k & 24\% & {0} (0)  \\ \hline
{cdn k256} & z48 & 487.1k & 162.4k & 89.7k & 428  & 1.1k & 0  & 637  & 0  &  69\%  & {18} (12)  & 19.5k & 22\% & {0} (0)  \\ \hline
{fdns} & z48 & 673.3k & 228.7k & 88.5k & 551  & 4.8k & 0  & 4.0k & 0  &  29\%  & {18} (6)  & 4.2k & 5\% & {-8} (0)  \\ \hline
{dnsdb} & z48 & 281.3k & 93.8k & 87.9k & 662  & 6.4k & 4  & 5.0k & 1  &  49\%  & {21} (12)  & 5.5k & 6\% & {-7} (0)  \\ \hline
{6gen} & z48 & 4.3M & 1.4M & 69.4k & 2.0k & 6.5k & 40  & 4.9k & 25  &  16\%  & {32} (12)  & 3.5k & 5\% & {-19} (-3)  \\ \hline
{caida} & z64 & 314.7k & 105.2k & 60.9k & 398  & 6.3k & 11  & 4.8k & 5  &  26\%  & {31} (12)  & 428  & 1\% & {-20} (-4)  \\ \hline
{caida} & z48 & 234.3k & 78.3k & 57.7k & 92  & 6.1k & 2  & 4.7k & 1  &  25\%  & {29} (12)  & 370  & 1\% & {-20} (-3)  \\ \hline
{fiebig} & z64 & 2.8M & 1.0M & 54.2k & 9.6k & 3.3k & 57  & 2.8k & 49  &  24\%  & {17} (5)  & 1.3k & 2\% & {-15} (-3)  \\ \hline
{fiebig} & z48 & 270.3k & 102.2k & 22.1k & 63  & 2.7k & 0  & 2.3k & 0  &  11\%  & {17} (4)  & 177  & 1\% & {-17} (-2)  \\ \hline

  \end{tabular}
  \vspace{-3mm}
\end{table*}

%% file: subnet-discovery.tex

\section{Subnet Discovery}\label{sec:subnetdiscovery}

Having collected voluminous trace results, we turn our
attention to topological inference from the results.
We employ two techniques to infer subnet boundaries.
The first one is based on path divergence observed in traces from
prior work (with IPv4), while the second relies on
the ubiquitous delegating of /64 prefixes as the
most-specific subnets at the Internet's periphery and thus is IPv6-specific.


{\bf Path divergence-based discovery.} Inspired by Lee
\etal\cite{lee2016identifying}, we discover \vsix subnets based
on path divergence detected in the paths traversed in
our trace campaigns.  However, IPv6 requires different premises.
First, our goal is to infer heterogeneous \vsix prefixes by splitting
subnets, starting with known BGP prefixes, into smaller ones based on
divergent hops; whereas subnets are coalesced to identify homogeneous
\vfour prefixes in~\cite{lee2016identifying}. Next, there is no
canonical equivalent of {\tt /24} in \vsix space due to the freedom
afforded network operators by generous address allotments and 
many transition and feature-based address assignment options.

As in Lee's work, the keys to the technique are twofold. First,
traced paths from one or more vantages to two different target
addresses are compared to {\em (a)} identify a {\em significant}
converging subpath (a substring, composed of router hop addresses,
that is common to both traced paths) and {\em (b)} identify a
subsequent {\em significant} diverging subpath. By ``significant''
we mean we are willing to assume that the divergence, in context
of the convergence prior, indicates that the two addresses belong
to different subnets.  We call the converging subpath by the name
``last common subpath'' (LCS) and refer to the diverging path tails by
the name ``divergent suffixes'' (DS). The second key to the technique is
that, once two addresses are assumed to be in different prefixes,
we calculate those two addresses' ``discriminating prefix length''
(DPL), introduced in Section~\ref{sec:DPL}.  Given that we take them to be
in different subnets, we know that the first $n$ bits of each address,
where $n$ is the DPL,
must also be the first $n$ bits of each subnet's base
address and, therefore, the subnet's prefix length must
be at least $n$.


Our implementation, discover\-By\-Path\-Div, classifies sets
of IPv6 traced paths as divergent (or not) according to these
parameters:

\squishlist
\item The LCS must have minimum length: $c=2$.
\item The LCS must have $C$ hop(s) having an ASN matching the target's ASN: $C=1$.
\item Missing hop addresses are not allowed in the LCS.
\item The last hop's ASN must not match the vantage's ASN: $A=1$.
\item The DS must have minimum length: $s=1$.
\item The DS must have $S$ hop(s) whose ASN match the target's: $S=1$.
\item No DS can have zero length: $z=0$.
\item The target ASN for each path pair must match: $T=1$.
\squishend

In this way, our implementation (with these numerous parameters)
can be made restrictive about the paths it accepts as evidence of
significant path divergence and, therefore, conservative in subnet
discovery. We tested even more conservative parameter values, \eg a
higher DS minimum length $s=2$, but path divergences
ostensibly due to traffic engineering and load-balancing occurred
too near the last hop, as reported also with IPv4 in prior works.

\replaced[id=2]{There}{Unfortunately, there} are additional complications. For instance,
{\em (a)} IPv6 networks exist that use many ASNs simultaneously,
\eg one originating routes to the BGP prefix(es) covering router
addresses and another originating routes for the prefix(es) covering
their customer's (target) addresses. To avoid these failing to meet
our path and target ASN requirements, we augment the BGP information
with collections of ``equivalent'' ASNs, \added[id=2]{identified by operational experience and expert knowledge, \eg of ISP acquisitions or mergers,} considering them equal even
though they are distinct numbers.  Also, {\em (b)} since it is not
necessary for networks to globally advertise routes to prefixes
covering their routers' addresses (since routers only need to communicate
in LANs or point-to-point links to each other), IPv6 networks also
exist that use router addresses not covered in the BGP. To avoid having
this violate our path and target ASN requirements, we augment the
BGP information with some prefixes that are in Regional Internet
Registries but not the global BGP. This is especially important
for IPv6 where a small number of very large networks, \eg
Comcast and Charter Spectrum, respectively, present these ASN and IP
prefix record keeping challenges.

{\bf /64 Discovery.} Our second subnet-related topology discovery
technique is a simpler one, specific to the IPv6 Internet which
typically has /64 subnets at its edge. This scheme is very
common since it is required by popular IPv6 address assignment
techniques such as SLAAC (Stateless Address Auto-configuration)
and SLAAC with temporary privacy extensions (also known as
``privacy addresses'') in which every IPv6 host address has an
interface identifier (IID) comprising the low 64 bits and a network
(subnet) identifier comprising the high 64 bits.

It is often the case that TTL-limited probes to some target address
never elicit an ICMP or ICMPv6 response with the target address
as its source.  This leaves the analyst not knowing whether
the resulting path reached the router nearest the target
or not, making it hard to make assertions on even simple metrics
like the diameter of the Internet (as measured in router hops).
To overcome this problem of \replaced{determining}{determine} where exactly that last hop
is in the topology, we leverage the fact, learned from prior active
measurement studies, that many IPv6 routers have the IID value {\tt
:0000:0000:0000:0001} (canonically displayed as {\tt ::1} with zero
compression) in the source address used to generate ICMPv6 error
messages such as Time-Exceeded~\cite{gray2015}, \eg routers serving as the gateway
for hosts having address in the subnet for that LAN.  Certainly not
all IPv6 routers do this, but it is common and is not unlike the
practice with IPv4 of, \eg using {\tt 192.168.0.1} as
the router gateway address for the {\tt 192.168.0.0/24} subnet.
When we see a last hop ending in {\tt ::1} that has the high 64-bits
matching the target address, we assume that the probe elicited a
response from the \added{gateway} router for the target's LAN, and thus completed.
This is a valuable inference in at least two ways: {\em (a)} it
allows us to assume this trace reached a unique /64 subnet and thus the
target address must be in a different subnet than another target
address that, likewise, has a last hop \replaced{in a different}{it its own} /64 covering
prefix, thereby enabling divergence-based discovery; {\em (b)} it
can help us infer reachability, access control, and firewalling
policies \replaced{that}{and} limit results in active measurement campaigns, \eg to assess
vulnerabilities or to discover topological characteristics.

We added this second technique to discover\-By\-Path\-Div
and call it the ``Identity Association (IA) Hack'' because
it can be leveraged to reverse-engineer the IP address
identity (prefix) delegated to a customer by an ISP. This
is is an important capability for privacy-minded Internet
operation if we wish to guarantee some level of anonymity in IP
addresses~\cite{DBLP:journals/corr/PlonkaB17}; we are pursuing this
as future work.


%% file: subnet.tex


{\bf Candidate Subnet Results.} We refer to our results
as ``candidates'' because our method determines a lower bound
on a subnet's prefix length. This bound is set based on
the highest DPL of a target address ostensibly within the subnet
and another ostensibly without, \ie where the traces
show paths to those targets diverged.  Thus, a candidate subnet
length means that we've discovered a subnet having a prefix length
of {\em at least} that reported.

\begin{figure}[tb]
\begin{centering}
  \begin{subfigure}{0.42\textwidth}
  \centering
  \includegraphics[width=\linewidth]{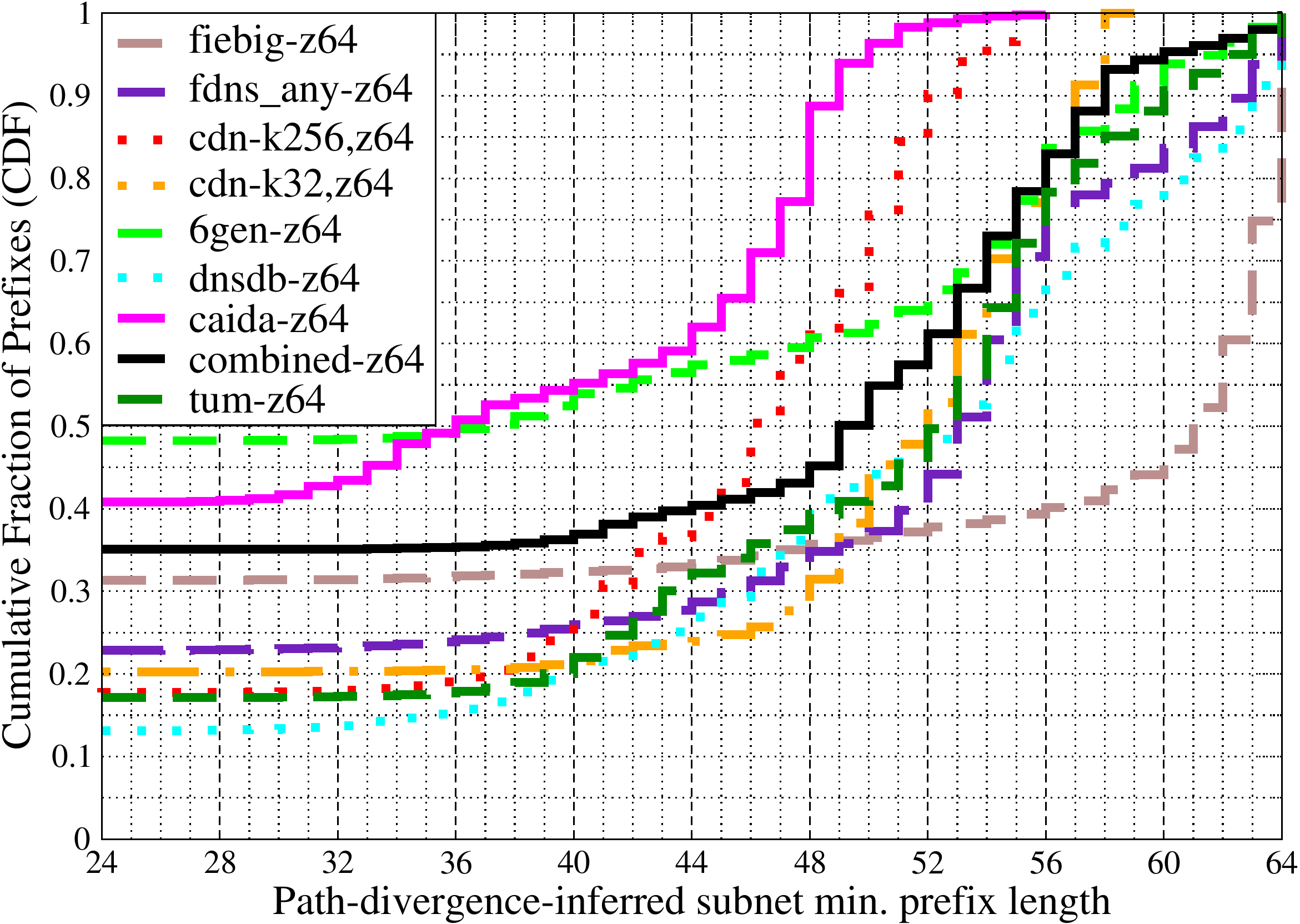}
  \caption{Distribution of minimum subnet prefix lengths}
  \label{fig:discCDF}
  \end{subfigure}
  \begin{subfigure}{0.42\textwidth}
  \includegraphics[width=\linewidth]{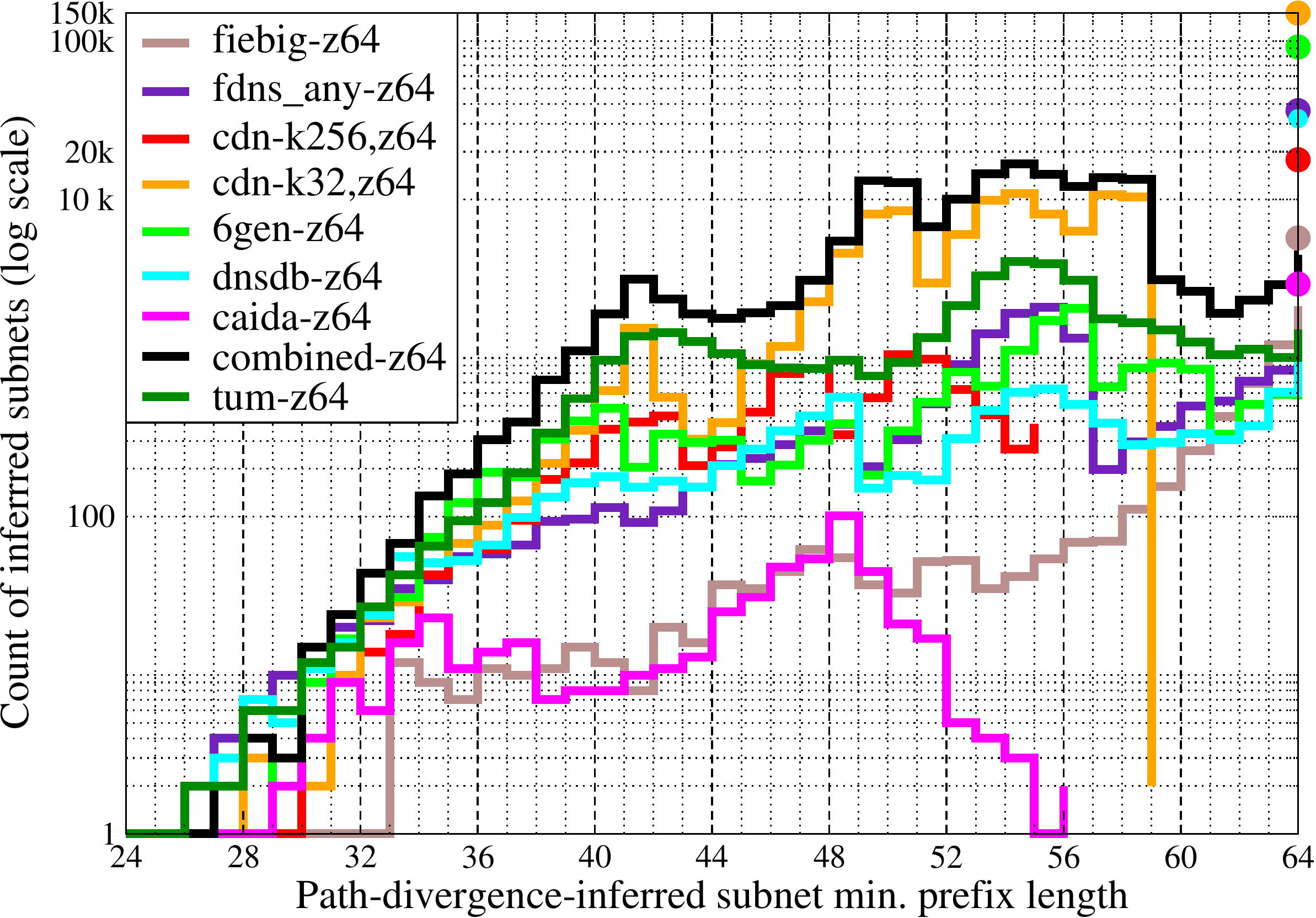}
  \caption{Counts of subnets by prefix length}
  \label{fig:discCounts}
  \end{subfigure}
  \vspace{-3mm}
  \caption{Subnets inferred by path divergence}
  \vspace{-5mm}
  \label{fig:discoverByPathDiv}
\end{centering}
\end{figure}


Examining the combination of all 45.8M traces for
path divergences, we \deleted[id=2]{find }discovered 172,497 candidate subnets,
covered by 1,726 BGP prefixes having 1,013 origin ASNs.
Figure~\ref{fig:discCDF} plots the CDF
of those subnets for each target set.
This figure shows that target sets power to discover candidate
subnets is largely governed by their respective addresses' DPLs
plotted in Figure~\ref{fig:targDPL}. Also note that, although we see
improved {\em potential} power of the target\deleted[id=2]{s} sets in combination
(Figure~\ref{fig:targCombinedDPL}), that improved power is not realized
here: the CDFs do not shift right.
This may be due to active measurement difficulties,
\eg missing hops on traced paths and our conservative
approach which does not allow missing hops on the common subpath
before divergence.

Figure~\ref{fig:discCounts} plots counts of discovered subnets
by prefix length per target set and combined.  Note there are a
few prominences suggesting popular prefix lengths (subnet sizes).
Also note that, while some target sets help to discover
only subnets less than a particular length, \eg 59 ({\tt cdn-k32})
or 55 ({\tt cdn-k256}), others can help to discover more specific
subnets because they contain unaggregated, active /64 prefixes, \eg from
the DNS.  The dots plotted directly above 64 on the horizontal axis
in Figure~\ref{fig:discCounts} are the counts \replaced{where}{times} the last hop
address was covered by the same /64 prefix as the target address,
\ie the IA Hack. (Combined these total to 1,284,891.)

{\bf Subnet Validation.}
To evaluate these results, first we use \deleted{as set of }ground truth data
consisting of a set of 12,447 interior prefixes of major US ISP
networks (109 BGP prefixes advertised by 30 origin ASNs) and their
respective city-level locations and assume those in
different cities are
also topologically heterogeneous, \ie in different subnets such
that \deleted{our }our method should be able to ascertain, if it is given
sufficient traces to target addresses in and near those subnets.



We find, that we've performed 386,579 traces from each vantage to
target addresses covered by 5,839 (47\%) of the subnets in truth
data.  Of those, our algorithm discovered 109 subnets exactly as in
the truth data, \ie having the same base addresses \added{and} prefix
lengths: 107 /40 prefixes and 2 /64 prefixes. However, because
these ground truth subnets are intermediate (in between advertised
BGP prefixes and LAN subnets) or ``distribution'' subnets,
we shouldn't expect many exact matches as our
method may have discovered more-specific subnets. Indeed we find that we've
discovered more specific candidates within 3,871 (66\%) of those ground
truth prefixes.


To deal with this complication,
we re-run our algorithm on a subset of traces selected by stratified
sampling, \ie we choose only one trace to one target address in each
ground truth subnet. This intentionally reduces
the fidelity of our technique by lowering the target addresses' DPLs,
thus limiting discovery to subnets no
more-specific than those in the truth data.
With this sampled subset of traces our algorithm yields 914
candidate subnets (18\%), 395 (43\%) of
which exactly match the truth data.  Of the non-matching results,
52\% have a prefix length that is one bit short and 20\% are short by two.
Improving this prefix length approximation by lower bound necessitates
additional traces to targets with higher DPL.  As such, we merely claim these
discovered subnet results are plausible based on modest
coincidence with ground truth and that stratified sampling
is a possible way the technique can be tuned to discover hierarchical
subnets.

\added{Attempting to further evaluate subnet discovery results, we examine
two universities' subnet address plans that were shared publicly.
For both universities,
the entirety of each of their active IPv6 address spaces is covered
by just one subnet in our results,
having prefix lengths of 41 and 39, respectively.
Investigation shows that
the anonymization method employed~\cite{DBLP:journals/corr/PlonkaB17}, \eg for
{\tt cdn-k32}, kept us from probing targets in these networks
ostensibly having too few simultaneously active IPv6 WWW clients.
That is, for privacy reasons, no target addresses with sufficiently high DPL
were probed to discern more-specific subnets.}

%% file: discuss.tex

\section{Discussion and Future Work}
\label{sec:discuss}


In this work, we seek to advance the state-of-the-art in Internet-wide \vsix
active topology mapping. We collect seeds and synthesize targets from a number
of sources using a three-step process. Next, to facilitate large-scale \vsix
topology mapping, we emit traceroutes from three vantage points to the targets
using a modified version of \yrp (\yrpsix). Along the way, we investigate the use of
various target selection methods and parameters (\eg maximum TTL,
probing speed, etc.) that elicit the most \vsix topological information. Our
investigations provide breadth across networks, depth to discover subnetting,
and speed. More specifically, we discover over 1.3M \vsix interface addresses,
which is an order of magnitude more compared to the state-of-the-art mapping
systems.

\subsection{Security~\added{\& Privacy}}
\label{sec:discuss:security}

Our results led us to consider 
security concerns specific to \vsix topology mapping. 
Due in part to the address length, \vsix addresses can contain sensitive information that makes
the active \vsix address space easier to scan or probe, and likely
more vulnerable to malicious exploits~\cite{rfc7707}. 
For example,
our probe campaigns \replaced{elicited}{generated} \icmpsix ``Time Exceeded'' responses
from many sources having \added{IEEE MAC-based} EUI-64 addresses.
These IIDs ostensibly embed Ethernet
addresses, exposing the manufacturer or model
of a router~\cite{acsac16furious}~\added{, and, perhaps, its operating system.
RFC 7721~\cite{rfc7721} notes that these IEEE-identifier-based IIDs
make attacks possible, \eg address scanning and device exploits}.
Likewise, we received ``Time Exceeded'' messages
from many addresses covered
by the same /64 prefix as a target address\replaced{, \ie routers
ostensibly collocated, \eg in a LAN, with hosts.
The community should
carefully consider the implications of these router-addressing practices
as a router's source address, \eg when sending \icmpsix error messages,}
{The community should
more carefully consider the implications of this address /64 co-location as the
source address, alone,}
can disclose details that users may prefer to remain private.


Therefore, we \deleted{chose to }release \deleted{two \vsix topology }datasets with
\deleted{different }restrictions.  The public~\replaced{data}{ly available topology}, 
available at \cite{topodownload}, \replaced{contains seeds and targets}{removes hops containing EUI-64 addresses
addresses, as 
well as those addresses
covered by the same /64 prefix as a target address}.  The complete
\replaced{data and results }{topology }will be available to researchers at \cite{dhs-impact} with restricted
distribution.
However, we note that our methodology is reproducible using
publicly-available
seed datasets and freely-available probe utilities.
The resulting addresses and subnets discovered or inferred
likely extend the attack surface already \replaced{discernible via }{provided by the }hitlists\deleted{ to routing infrastructure}.
%


\subsection{Future Work}
\label{sec:future}

Based on these encouraging results, we plan to leverage our
methodology across a large number of vantages and time to 
provide even greater scope and coverage.  Further, we plan to
perform alias resolution, \eg\cite{imc13speedtrap}, to produce
router-level topologies and facilitate
comparative graph analyses between \vfour and \vsix.

Finally, our subnet discovery results show that it is sometimes feasible to
remotely determine likely \vsix prefix assigned to
a single user or subscriber.
In DHCPv6 prefix delegation this is referred to as
Identity Association (IA), and prefix lengths can vary according to
the services offered and hints offered by a router to
which a prefix is delegated, \eg on the customer premises.
We plan to run additional measurement campaigns to comprehensively
assess this capability, with the hope that it might inform IP address
anonymization by aggregation, to provide a known level of client privacy.
